\documentclass[12pt]{JHEP3}
\usepackage{axodraw}

\def\Year{\expandafter\eatPrefix\the\year}
\newcount\hours \newcount\minutes
\def\monthname{\ifcase\month\or
January\or February\or March\or April\or May\or June\or July\or
August\or September\or October\or November\or December\fi}
\def\shortmonthname{\ifcase\month\or
Jan\or Feb\or Mar\or Apr\or May\or Jun\or Jul\or Aug\or Sep\or
Oct\or Nov\or Dec\fi}

\def\TimeStamp{\hours\the\time\divide\hours by60%
\minutes -\the\time\divide\minutes by60\multiply\minutes by60%
\advance\minutes by\the\time%
${\rm \shortmonthname}\cdot   \if\day<10{}0\fi\the\day\cdot
\the\year \qquad\the\hours:\if\minutes<10{}0\fi\the\minutes$}

\newskip\humongous \humongous=0pt plus 1000pt minus 100pt
\def\caja{\mathsurround=0pt}
\def\eqalign#1{\,\vcenter{\openup1\jot \caja
       \ialign{\strut \hfil$\displaystyle{##}$&$
        \displaystyle{{}##}$\hfil\crcr#1\crcr}}\,}
\newif\ifdtup

\newcounter{eqnumber}[section]
\renewcommand{\theeqnumber}{\thesection.\arabic{eqnumber}}
\def\equn{\refstepcounter{eqnumber}
\eqno({\rm \theeqnumber}) }

\def\eqn#1{eq.~(\ref{#1})}

\def\eqns#1#2{eqs.~(\ref{#1}) and~(\ref{#2})}

\def\fig#1{fig.~{\ref{#1}}}

\def\sec#1{section~{\ref{#1}}}

\def\tree{{\rm tree}}

\def\cc{\dagger}

\newbox\charbox
\newbox\slabox
\def\s#1{{      
        \setbox\charbox=\hbox{$#1$}
        \setbox\slabox=\hbox{$/$}
        \dimen\charbox=\ht\slabox
        \advance\dimen\charbox by -\dp\slabox
        \advance\dimen\charbox by -\ht\charbox
        \advance\dimen\charbox by \dp\charbox
        \divide\dimen\charbox by 2
        \raise-\dimen\charbox\hbox to \wd\charbox{\hss/\hss}
        \llap{$#1$}
}}

\def\spa#1.#2{\left\langle#1\,#2\right\rangle}
\def\spb#1.#2{\left[#1\,#2\right]}
\def\lor#1.#2{\left(#1\,#2\right)}
\def\sign{{\mathop{\rm sign}\nolimits}}

\catcode`@=11  

\def\Tr{\, {\rm Tr}}

\def\eps{\epsilon}

\def\e{\epsilon}

\def\half{{1\over 2}}

\def\la{\langle}
\def\ra{\rangle}
\def\oneloop{{1 \mbox{-} \rm loop}}

\def\lsl{\not{\hbox{\kern-2.3pt $\ell$}}}
\def\Psl{\not{\hbox{\kern-2.3pt $P$}}}
\def\ksl{\not{\hbox{\kern-2.3pt $k$}}}
\def\twosl{\not{\hbox{\kern-2.3pt $2$}}}
\def\fivesl{\not{\hbox{\kern-2.3pt $5$}}}

\def\rg{r_{\Gamma}}

\def\spa#1.#2{\left\langle#1\,#2\right\rangle}
\def\spb#1.#2{\left[#1\,#2\right]}
\def\lor#1.#2{\left(#1\,#2\right)}

\def\sand#1.#2.#3{%
  \left\langle\smash{#1}{\vphantom1}\right|{#2}%
  \left|\smash{#3}{\vphantom1}\right\rangle}
\def\sandp#1.#2.#3{%
  \left\langle\smash{#1}{\vphantom1}^{-}\right|{#2}%
  \left|\smash{#3}{\vphantom1}^{+}\right\rangle}
\def\sandpp#1.#2.#3{%
  \left\langle\smash{#1}{\vphantom1}^{+}\right|{#2}%
  \left|\smash{#3}{\vphantom1}^{+}\right\rangle}
\def\sandmm#1.#2.#3{%
  \left\langle\smash{#1}{\vphantom1}^{-}\right|{#2}%
  \left|\smash{#3}{\vphantom1}^{-}\right\rangle}
\def\sandpm#1.#2.#3{%
  \left\langle\smash{#1}{\vphantom1}^{+}\right|{#2}%
  \left|\smash{#3}{\vphantom1}^{-}\right\rangle}
\def\sandmp#1.#2.#3{%
  \left\langle\smash{#1}{\vphantom1}^{-}\right|{#2}%
  \left|\smash{#3}{\vphantom1}^{+}\right\rangle}

\def\BR#1#2{\la#1^+|K|#2^+\ra}
\def\Aloop{A^{\rm 1\hbox{-}loop}}

\def\Atree{A^{\rm tree}}

\def\dlips{dLIPS}
\def\NeqEight{{\cal N} = 8}
\def\NeqFour{{\cal N} = 4}
\def\NeqOne{{\cal N} = 1}

\def\Fn{n}
\def\Fs#1#2{F^{{#1}}_{\Fn:#2}}
\def\Fone{\Fs{\rm 1m}}
\def\Feasy{\Fs{{\rm 2m}\,e}}
\def\Fhard{\Fs{{\rm 2m}\,h}}
\def\Fthree{\Fs{\rm 3m}}
\def\Ffour{\Fs{\rm 4m}}

\def\tn#1#2{t^{[#1]}_{#2}}

\def\dlips{d{\rm LIPS}}

\def\cg{\hat{r}_\Gamma}
\def\Fact{{\cal F}}

\title{Inherited Twistor-Space Structure of Gravity Loop Amplitudes}
\author{Zvi~Bern\\ Department of Physics\\ University of California
at Los Angeles \\ Los Angeles,
CA 90095, USA}
\author{N.~E.~J.~Bjerrum-Bohr, David~C.~Dunbar\\ Department of Physics \\
University of Wales Swansea \\
Swansea, SA2 8PP, UK}

\preprint{SWAT-05-424 \cr
\hfill UCLA/05/TEP/1}

\abstract{ At tree-level, gravity amplitudes are obtainable directly
from gauge theory amplitudes via the Kawai, Lewellen and Tye
closed-open string relations.  We explain how the unitarity method
allows us to use these relations to obtain coefficients of box
integrals appearing in one-loop $\NeqEight$ supergravity amplitudes
from the recent computation of the coefficients for $\NeqFour$
super-Yang-Mills non-maximally-helicity-violating amplitudes.  We
argue from factorisation that these box coefficients determine the
one-loop $\NeqEight$ supergravity amplitudes, although this remains to
be proven.  We also show that twistor-space properties of the
$\NeqEight$ supergravity amplitudes are inherited from the corresponding
properties of $\NeqFour$ super-Yang-Mills theory.  We give a number of
examples illustrating these ideas.}

\begin{document}

\section{Introduction}

Scattering amplitudes in gravity theories are closely related to those
of gauge theory. At tree level there exists a set of general relations
expressing gravity tree amplitudes as sums of products of gauge theory
ones.  These relations follow from the low energy limit of the
Kawai-Lewellen-Tye (KLT) relations between open and closed string
theory amplitudes~\cite{KLT,BerGiKu,GravityReview}.  In this limit the string
relations reduces to relations for effective field theories of
gravity~\cite{Weinberg:1978kz,Donoghue:1994dn}.
Moreover, in the low energy limit these relations do not
require the existence of an underlying consistent string theory and
hold in any dimensions or massless matter
contents~\cite{BDWGravity}. The relations also hold for large classes
of higher dimension operators in the effective field
theory~\cite{EffKLT}.

At loop level, the standard methods for constructing  amplitudes
via Feynman rules provide no obvious means of exploiting the KLT
relations. An alternative is provided by the unitarity method of
Dixon, Kosower and two of the
authors~\cite{BDDKa,BDDKb,BernMorgan,TwoloopSplitting}.
This method is ideal for exploiting the KLT
relations at loop level, since the method obtains loop amplitudes
directly from tree amplitudes, which do satisfy the KLT relations.
More generally when coupled with the KLT relations, the unitarity
method allows advances in gauge theory loop calculations to be carried
over to gravity calculations.  These ideas have been used to compute
the MHV amplitudes of pure gravity~\cite{Bern:1998xc} and $\NeqEight$
supergravity~\cite{Bern:1998sv} and to demonstrate
that $\NeqEight$ supergravity~\cite{ExtendedSugra} is less divergent
in the ultraviolet than had been expected~\cite{BDDPR,Howe:2002ui}
previously. Recently, there have been significant advances in gauge
theory computations, stimulated by Witten's proposal of a
twistor-space topological string
theory~\cite{WittenTopologicalString,RSV,Berkovits,BerkovitsMotl,Vafa,Siegel:2004dj}
as a candidate for a weak--weak duality to maximally supersymmetric
gauge theory.  This string theory generalises Nair's earlier
description~\cite{Nair} of the simplest gauge-theory amplitudes.  The
twistor-space structure of the amplitudes implies that gauge theory
amplitudes are simpler than had been suspected previously.  In
particular, for massless gauge theories, Cachazo, Svr\v{c}ek and
Witten (CSW)~\cite{Cachazo:2004kj} have presented a set of new
computational rules for scattering amplitudes, in terms of ``MHV
vertices'' inspired by the twistor space structure. Other simple
versions of tree amplitudes may be found in
ref.~\cite{CSW:matter,Tree,BDDK7, BDKn,Roiban:2004ix,BrittoUnitarity}.

At loop level, a direct topological-string approach appears to be problematic
because of the appearance of non-unitary states from
conformal supergravity~\cite{BerkovitsWitten}.  Nevertheless,
significant progress has been accomplished at loop level by more
direct means.  An important step, clarifying the structure of loop
amplitude, is the computation by Brandhuber, Spence and
Travaglini~\cite{Brandhuber:2004yw} of the $\NeqFour$ MHV amplitudes
using MHV vertices. Since then there has been rapid progress in
obtaining amplitudes in $\NeqFour$ and $\NeqOne$
theories~\cite{Cachazo:2004zb,Cachazo:2004by,BBKR,Cachazo:2004dr,
Britto:2004nj,BDDK7,Britto:2004tx,BBDD,BBDP,Quigley:2004pw,
Bedford:2004py,Bedford:2004nh},
including the recent computation of all $\NeqFour$ next-to-MHV (NMHV)
one-loop amplitudes~\cite{BDKn} and the next-to-next-to-MHV (N$^2$MHV)
box coefficient~\cite{BrittoUnitarity}.  These calculations rely on the
four-dimensional cut constructibility of the
amplitudes~\cite{BDDKa,BDDKb} and on the knowledge of having a basis of
dimensionally regularised integrals~\cite{Integrals5}. An important
recent development, enhancing the power of the unitarity method, is
the observation by Britto, Cachazo and Feng~\cite{BrittoUnitarity} that
box integral coefficients can be obtained from generalised unitarity
cuts~\cite{Eden,GeneralizedCuts,TwoloopSplitting,BDDK7} by solving the
on-shellness constraints in signature $(--++)$.

The situation with gravity is less clear.  As explained in Witten's
original paper~\cite{WittenTopologicalString}, gravity amplitudes will have
a derivative of a $\delta$-function support rather than a simple
$\delta$-function support on degenerate curves in twistor space.
This apparently
complicates their structure, preventing the construction of MHV
vertices for gravity amplitudes~\cite{GBgravity,ZhuGrav}.
Nevertheless, the KLT relations provide direct means for obtaining
gravity tree amplitudes from any newly computed gauge theory tree amplitudes.

In this paper, we follow the logic of
refs.~\cite{BDDPR,Bern:1998xc,Bern:1998sv}, and use the unitarity
method to carry over the recent gauge theory advances to quantum
gravity. At one loop, supergravity is ultraviolet finite, so we do not
have to concern ourselves with issues of non-renormalisability.
In ref.~\cite{BDDPR} Dixon, Perelstein, Rozowsky and two of
the authors observed that for four-point $\NeqEight$ supergravity, 
the coefficients of integral functions are
proportional to products of gauge theory integral coefficients.  In
this paper we generalise this notion to the box
coefficient of any one-loop gravity amplitude and show that in general
the box integral coefficients can be expressed in terms of products of
box coefficients appearing in the gauge theory.  This relation between
box integral coefficients holds in any theory (assuming we are working
in a basis of only $D=4-2\e$ integral functions) and does not rely on
the four-dimensional cut constructibility of the theory.  The case we
consider here explicitly is $\NeqEight$ supergravity.  The MHV amplitudes
of this theory have been worked out in ref.~\cite{Bern:1998sv}.

Although the $\NeqEight$ theory does not appear to satisfy the power
counting criterion required for four-dimensional cut constructibility,
we argue that the four-dimensional cuts determine the one-loop
amplitudes, as suggested by their factorisation properties.  Moreover,
we expect that one-loop $\NeqEight$ amplitudes are composed solely of
the box integral functions.  This may seem rather surprising, given
the apparent violation of the cut-constructibility power-counting
criterion.  However, in ref.~\cite{Bern:1998sv}, explicit computations
proved that up to six points in the MHV case, this miracle indeed
happens.  Furthermore, as argued in that reference, the factorisation
properties suggests that these miracles should continue as the number
of external legs increases.  Here we apply the same logic to argue
that even for non-MHV amplitudes we do not expect triangle, bubble or
additional rational functions to appear in any $\NeqEight$ one-loop
amplitude.  If true, with this ansatz we can obtain full $\NeqEight$
one-loop amplitudes rather directly from corresponding $\NeqFour$
super-Yang-Mills amplitudes, by relating the $\NeqEight$ supergravity
box coefficients to the $\NeqFour$ super-Yang-Mills box coefficients.

We also use the explicitly computed results to explore the
twistor-space structure of gravity amplitudes at both tree and
loop level. Not too surprisingly, given the close relationship of
gravity and gauge theory amplitudes, we find that the gauge theory
twistor properties induce closely corresponding twistor properties on
gravity amplitudes. The key difference, as already observed by Witten,
is instead of $\delta$-function support, twistor-space gravity amplitudes
have a derivative of a $\delta$-function support.  At tree-level the twistor
properties follow directly from analysing the amplitudes obtained via
the KLT relations.  At loop level we combine the unitarity method with
the KLT relations in order to deduce the twistor-space properties of
the coefficients of box integrals in gravity theories.

\section{The Kawai-Lewellen-Tye Relations }

Gravity amplitudes can be constructed through the KLT-relations
which connects the amplitudes for closed and open strings. The
general $n$-point scattering amplitude for a closed string is
connected to that of the open string through the following
formula~\cite{KLT}:
$$
{\cal M}_n^{\rm (closed\ string)} \sim \sum_{\Pi,\tilde \Pi}
e^{i\pi\Phi(\Pi,\tilde\Pi)} {\cal A}_n^{\rm left\ (open\ string)}(\Pi)
{\cal A}_n^{\rm right\ (open\ string)}(\tilde \Pi),
\equn
$$
where $\Pi(1,2,\ldots,n)$ and $\tilde\Pi(1,2,\ldots,n)$ are sets of
the external lines of the open string modes associated with particular
cyclic orderings. At infinite string tension, ($\alpha'
\longrightarrow 0$) the KLT-relationship relates the field theory tree
amplitudes of Yang-Mills theory and gravity. In ref.~\cite{EffKLT} this was
extended also to higher derivative effective field theories of gravity.

The explicit form of the KLT-relationship
up to six points at $\alpha' =0$ is,
$$
\eqalign{\hspace{2cm}
M_3^{\rm tree}(1,2,3) =
&
-iA_3^{\rm tree}(1,2,3)A_3^{\rm tree}(1,2,3),
\cr
M_4^{\rm tree}(1,2,3,4) =
&
-is_{12}A_4^{\rm tree}(1,2,3,4)A_4^{\rm tree}(1,2,4,3), \label{KLTFour} \cr
M_5^{\rm tree}(1,2,3,4,5) =&
\; is_{12}s_{34}\ A_5^{\rm tree}(1,2,3,4,5)A_5^{\rm tree}(2,1,4,3,5) \cr
& \hskip 1.9 cm \null
+i s_{13}s_{24}\ A_5^{\rm tree}(1,3,2,4,5)A_5^{\rm tree}(3,1,4,2,5),
\label{KLTFive} \cr
M_6^{\rm tree}(1,2,3,4,5,6) = &
-is_{12}s_{45}\ A_6^{\rm tree}(1,2,3,4,5,6)(s_{35}A_6^{\rm tree}(2,1,5,3,4,6)
\cr
& \hskip 1.9 cm \null
+(s_{34}+s_{35})\ A_6^{\rm tree}(2,1,5,4,3,6)) + {\cal P}(2,3,4)\,,
\cr}
\equn\label{KLTSix}
$$
where
$s_{ij} = (k_i + k_j)^2$, ${\cal P}(2,3,4)$ represents the sum over
permutations of legs $2,3,4$ and the $A^\tree_n$ are tree-level
colour-ordered gauge theory partial
amplitudes~\cite{TreeReview,TreeColour,Colour}.  The complete gauge
theory amplitudes are obtained by multiplying these by the colour
structures and by summing over permutations.  In general throughout
the paper, gauge theory amplitudes will be denoted by $A$ and gravity
ones by $M$. The KLT relations have been explicitly presented for an
arbitrary number of legs~\cite{Bern:1998sv} and these combine to give
the full amplitudes via,
$$
\eqalign{
{\cal M}_n^{\rm tree}(1,2,\ldots, n) &=
\left({  \kappa \over 2} \right)^{(n-2)}
M_n^{\rm tree}(1,2,\ldots, n)\,,
\cr
{\cal A}_n^{\rm tree}(1,2,\ldots, n) &= g^{(n-2)} \hspace{-0.1cm}\sum_{\sigma \in S_n/Z_n}
\hspace{-0.2cm}{\rm Tr}\left( T^{a_{\sigma(1)}}
T^{a_{\sigma(2)} }\cdots  T^{a_{\sigma(n)}} \right)
 A_n^{\rm tree}(\sigma(1), \sigma(2),\ldots, \sigma(n)) \, ,
\cr}
\equn
$$
where $S_n/Z_n$ is the set of all permutations, but with cyclic
rotations removed. The $T^{a_i}$ are fundamental representation
matrices for the Yang-Mills gauge group $SU(N_c)$, normalised so that
$\Tr(T^aT^b) = \delta^{ab}$.
(For more detail on the tree and one-loop colour ordering of
gauge theory amplitudes see refs.~\cite{TreeReview,Colour}.)

\section{Unitarity Method}

In the unitarity-based method, loop amplitudes are constructed
from tree amplitudes by considering the various cases where
internal propagators go on shell.
Letting two propagators go on
shell is equivalent to evaluating a phase space integral over
products of tree amplitudes,
$$
\eqalign{
C_{i,\ldots,j}   & \equiv\hspace{-0.1cm}
{ i \over 2}\hspace{-0.1cm} 
\int \dlips\biggl[ A^{\rm tree}(\ell_1,i,i+1,\ldots,
j,\ell_2) 
A^{\rm tree}(-\ell_2,j+1,j+2,\ldots,i-1,-\ell_1)
\biggr]\;.
\cr}
\equn
$$
This phase space integral gives the discontinuity of the amplitude
in the cut channel.
In general, we may expand the amplitude as sum of dimensionally regularized
integral functions with rational
coefficients~\cite{BDDKa,BDDKb,Integrals5},
$$
 \Aloop= \sum_{a} \hat c_a I_a\,.
\equn
$$
We may then obtain the rational coefficients $c_a$
from the cuts of the one-loop amplitude
$$
 {\rm Im}_{K_{i,\ldots,j} > 0} A^{\rm 1\hbox{-}loop} = \sum_a \hat c_a
\; {\rm Im}_{K_{i,\ldots,j} > 0}( I_a)\,,
\equn $$
and their
generalizations~\cite{BDDKa,BDDKb,BernMorgan,TwoloopSplitting,
GeneralizedCuts,TwoloopSplitting,BDDK7,BrittoUnitarity}.

We will use two representations of the integral
functions. Firstly, the scalar box integrals, $I$, as defined in
ref.~\cite{Integrals5} and secondly the rescaled box-functions denoted $F$,
$$
I_4 = { 1 \over D} F\,,
\equn\label{IvsF}
$$
where $D$ is a kinematic denominator quadratic in momentum invariants.
Explicitly, from ref.~\cite{BDDKa},
$$
\eqalign{
  I_{4:i}^{1{\rm m}} &=\  {(-2 \rg)\Fone{i} \over \tn{2}{i-3} \tn{2}{i-2} }
        \,,  \hskip 0.5 cm
 I_{4:r;i}^{2{\rm m}e}
=\  {(-2 \rg)\Feasy{r;i}
      \over \tn{r+1}{i-1}\tn{r+1}{i} -\tn{r}{i}\tn{n-r-2}{i+r+1} }\,,
        \,  \hskip 0.5 cm
  I_{4:r;i}^{2{\rm m}h}
=\  {(-2 \rg)\Fhard{r;i} \over \tn{2}{i-2} \tn{r+1}{i-1} } \,,
\cr
  I_{4:r,r',i}^{3{\rm m}}
&=\  {(-2 \rg)\Fthree{r,r';i}
     \over \tn{r+1}{i-1} \tn{r+r'}i -\tn{r}{i} \tn{n-r-r'-1}{i+r+r'} }\,,
        \,  \hskip 0.5 cm
I_{4: r, r', r'', i}^{4{\rm m}}  =
 {(-2 \rg)\Ffour{r, r', r'';i}\over t_i^{[r+ r']}\; 
t_{i+r}^{[r'+r'']}\;\rho}\, ,
\cr}\equn\label{ExplicitIvsF}
$$
where we use the notation of that reference. In particular, $t_i^{[r]} =
(k_i+ \cdots + k_{i+r-1})^2$. (See the first appendix of
ref.~\cite{BDDKa} for definitions and a more detailed description of
the integral functions.)  We shall use $\hat c_a$ for
coefficients of $I$ and $c_a$ for coefficients of $F$. We can move
between the two representation using
$$ 
\hat c_a =   { c_a  D}\,,
\equn\label{hatcvsc}
$$
where the kinematic denominator $D$ for each type of integral may be read off 
by comparing \eqn{IvsF} and \eqn{ExplicitIvsF}.

One-loop massless amplitudes, which satisfy the power counting
criterion that $n$-point Feynman integrals have no more than $n-2$
powers of loop momenta in their numerators, can be obtained directly
from four-dimensional tree amplitudes~\cite{BDDKb}.  Hence, when
this criterion is satisfied, one may fix all rational functions
appearing in the amplitudes directly from the terms
which contain cuts in four dimensions.  We refer to such amplitudes as
``cut-constructible''. Supersymmetric gauge theory amplitudes, in
particular, satisfy this criterion and in the case of $\NeqFour$
super-Yang-Mills theory, the amplitude is expressible entirely as a
linear combination of box integral functions~\cite{BDDKa}.

\FIGURE[ht]{
\begin{picture}(100,100)(0,0)
\DashLine(50,73)(50,61){2}
\DashLine(50,39)(50,27){2}
\DashLine(27,50)(39,50){2}
\DashLine(61,50)(73,50){2}

\Line(30,30)(30,70)
\Line(70,30)(70,70)
\Line(30,30)(70,30)
\Line(70,70)(30,70)

\Line(30,70)(20,70)
\Line(30,70)(30,80)

\Line(70,30)(70,20)
\Line(70,30)(80,30)

\Line(70,70)(70,80)
\Line(70,70)(80,70)

\Line(30,20)(30,30)
\Line(20,30)(30,30)

\Text(30,10)[c]{$i_7$}
\Text(10,30)[c]{$i_8$}
\Text(25,25)[c]{$\bullet$}
\Text(70,10)[c]{$i_6$}
\Text(90,30)[c]{$i_5$}
\Text(70,90)[c]{$i_3$}
\Text(90,70)[c]{$i_4$}
\Text(30,90)[c]{$i_2$}
\Text(10,70)[c]{$i_1$}
\Text(75,25)[c]{$\bullet$}
\Text(75,75)[c]{$\bullet$}
\Text(25,75)[c]{$\bullet$}

\Text(20,52)[c]{$\ell_1$}
\Text(52,20)[c]{$\ell_4$}
\Text(80,52)[c]{$\ell_3$}
\Text(52,80)[c]{$\ell_2$}
\end{picture}
\label{QuadrupleCutFigure}
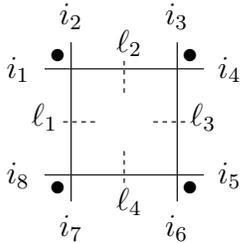
\caption{A quadruple cut of a $n$-point amplitude.  The dashed
lines represent the cuts.  The dot represents an arbitrary number of
external line insertions.}
}

A key property that allows us to relate the coefficients of box
integrals in the $\NeqFour$ super-Yang-Mills theory to those of
$\NeqEight$ supergravity is that the coefficients of the integrals can
be determined by purely algebraic means starting from the unitarity
cuts.  The integral reduction method of van~Neerven and
Vermaseren~\cite{vNV} is an example of a reduction formalism with the
property that in a given cut we may algebraically link the
coefficients of the integrals to the original expressions for the
cuts. Alternatively, it is more convenient to use generalised cuts
~\cite{Eden,GeneralizedCuts,TwoloopSplitting,BDDK7}, where multiple
propagators go on shell.  Using the recent
observation~\cite{BrittoUnitarity}, that box integral coefficients can
be directly obtained algebraically from generalised quadruple cuts one
can straightforwardly solve for the coefficients.  Since a given
generalised quadruple cut selects out a unique box integral function,
we may relate gravity and Yang-Mills coefficients via the KLT tree
level relations, since the cuts are expressed in terms of tree
amplitudes.  Specifically, if we consider an amplitude containing the
scalar box integral function shown in \fig{QuadrupleCutFigure}, then
the coefficient of this function is given by the product of the four
tree amplitudes where the cut legs fully satisfy on-shell 
conditions~\cite{BrittoUnitarity},
$$
\eqalign{
\hat c={ 1 \over 2 } \sum_{\cal S}
\biggl( \Atree(\ell_1,i_1,  & \ldots,i_2,\ell_2) \times
\Atree(\ell_2,i_3,\ldots,i_4,\ell_3)
\cr
& \hspace{2cm}\times \Atree(\ell_3,i_5,\ldots,i_6,\ell_4) \times
\Atree(\ell_4,i_7,\ldots,i_8,\ell_1) \biggr)\;,
\cr}
\equn
$$
where ${\cal S}$ indicates the set of helicity configurations of
the legs $\ell_i$ which give a non-vanishing product of tree
amplitudes.
Employing signature $(--++)$ or complex momenta is useful in this context
because it allows one to use this formula even when one of the tree
amplitudes is a three-point amplitude: In Minkowski signature with
real momenta such on-shell tree amplitudes vanish.

\section{Gauge Theory Results}

In this section we will collect various results in $\NeqFour$ gauge
theory that we use in \sec{GravityAmplitudesSection} to obtain gravity
one-loop amplitudes.

\subsection{MHV Amplitudes}

The $\NeqFour$ MHV amplitudes are remarkably simple and were first
calculated in ref.~\cite{BDDKa}. They have also been re-computed using
twistor-inspired methods~\cite{Brandhuber:2004yw,Cachazo:2004dr} and
are given by simple linear combinations of box integral
functions,
$$
 A_{n;1}^{\NeqFour\ {\rm MHV}} =
   \hat r_\Gamma \, A_{n}^{\rm tree} \times  V_n^g \,.
\equn\label{MHVAmplitudes}
$$
The factor $V_n^g$ ($n\ge 5$) depends on whether $n$ is odd
($n=2m+1$) or even ($n=2m$),
$$
\eqalign{
(\mu^2)^{-\eps} V_{2m+1}^g = &  \sum_{r=2}^{m-1} \sum_{i=1}^{n}
 F^{2{\rm m} \, e}(s_{i\ldots (i+r)}, s_{(i-1)\ldots (i+r-1)},
          s_{i \ldots (i+r-1)}, s_{(i+r+1) \ldots (i-2)})  \cr
& \hskip 5.8cm \null
     + \sum_{i=1}^{n} F^{1{\rm m}} (s_{i-3,i-2}, s_{i-2,i-1},
                s_{i \ldots (i-4)})
       \,, \cr
(\mu^2)^{-\eps} V_{2m}^g = &
\sum_{r=2}^{m-2} \sum_{i=1}^{n}
  F^{2{\rm m} \, e}(s_{i\ldots (i+r)}, s_{(i-1)\ldots (i+r-1)},
                   s_{i \ldots (i+r-1)}, s_{(i+r+1) \ldots (i-2)})  \cr
& \hskip 5.8 cm  \null
+ \sum_{i=1}^{n}  F^{1{\rm m}} (s_{i-3,i-2}, s_{i-2,i-1},
                s_{i \ldots (i-4)}) \nonumber \cr
& \hskip 1.3 cm  \null
+ \sum_{i=1}^{n/2} F^{2{\rm m} \, e}( s_{i \ldots (i+m-1)},
                                  s_{(i-1) \ldots (i+m-2)},
                                  s_{i \ldots (i+m-2)},
                                  s_{(i+m) \ldots (i-2)}) \,. \hskip 1 cm
\cr}
\equn\label{CoeffDefine}
$$
using the box integral functions $F$ as defined in eq.~(\ref{ExplicitIvsF})
and given in the first appendix of
ref.~\cite{BDDKa}. In the above
$$
\cg\ =\ {1 \over (4 \pi)^{2-\e}} r_\Gamma =  {1 \over (4 \pi)^{2-\e}}
{\Gamma(1+\e)\Gamma^2(1-\e)\over\Gamma(1-2\e)}\,,
\equn\label{cGamma}
$$
is a  prefactor. Because
of the dimensional regulator, an overall factor of $(\mu^2)^\e$
enters where $\mu$ is an arbitrary scale.

\subsection{NMHV Amplitudes}
\label{6ptYMSubsection}

First we consider the six-gluon amplitudes of $\NeqFour$ super-Yang-Mills
theory.   These amplitudes
were calculated in ref.~\cite{BDDKb}.  We collect these results
here, since we will use them in \sec{GravityAmplitudesSection} to
obtain the corresponding $\NeqEight$ supergravity amplitudes.
There are three independent NMHV super-Yang-Mills
partial amplitudes.  Since the leading colour
gauge theory amplitudes are colour-ordered
we need only consider one cyclic ordering.  (Subleading colour partial
amplitudes may be obtained from the leading ones, by summing over
appropriate permutations~\cite{BDDKa}).

In general, the six-point box coefficients are of the form~\cite{BDDKb},
$$
c_{} = c_{NS} + c_S \,,
\equn\label{SingletPlusNonSinglet}
$$ 
where the two terms arise from different helicity structures in the
cuts three-particle channels, as illustrated in
\fig{SingletNonsingletFigure}.  (In one case below one of the two
terms vanishes.) A six-point box has only one of its cuts in a
three-particle channel.  The cut in the three particle channel may be
divided into ``singlet'' and ``non-singlet'' contributions as shown in
\fig{SingletNonsingletFigure}. The coefficient $c_{NS}$ represents the
non-singlet contribution and $c_{S}$ the singlet contribution.  The
singlet term corresponds to the two cut legs having the same helicity
on one side of the cut. The singlet terms thus has contributions only
from gluons crossing the cut. The non-singlet term has its cut legs
having opposite helicity on one side of the cut.  For this
configuration all terms in the $\NeqFour$ multiplet contribute.

\FIGURE[h]{
\begin{picture}(200,110)(-100,0)
\DashLine(27,50)(39,50){2}
\DashLine(61,50)(73,50){2}

\Line(30,30)(30,70)
\Line(70,30)(70,70)
\Line(30,30)(70,30)
\Line(70,70)(30,70)

\Line(30,30)(20,20)
\Line(70,30)(80,20)

\Line(30,70)(20,70)
\Line(30,70)(30,80)

\Line(70,70)(70,80)
\Line(70,70)(80,70)

\Text(13,15)[l]{$3^-$}
\Text(78,15)[l]{$4^+$}
\Text(7,72)[l]{$5^+$}
\Text(26,88)[l]{$6^+$}

\Text(66,88)[l]{$1^-$}
\Text(83,72)[l]{$2^-$}

\Text(-90,50)[l]{\bf NON-SINGLET:}

\Text(35,40)[c]{$^+$}
\Text(35,55)[c]{$^-$}
\Text(42,64)[c]{$^-$}
\Text(60,64)[c]{$^+$}
\Text(66,55)[c]{$^+$}
\Text(66,40)[c]{$^-$}
\Text(60,33)[c]{$^-$}
\Text(42,33)[c]{$^+$}

\SetWidth{1.5}
\DashLine(50,20)(50,80){3}
\end{picture}
\begin{picture}(220,110)(-75,0)
\DashLine(27,50)(39,50){2}
\DashLine(61,50)(73,50){2}

\Line(30,30)(30,70)
\Line(70,30)(70,70)
\Line(30,30)(70,30)
\Line(70,70)(30,70)

\Line(30,30)(20,20)
\Line(70,30)(80,20)

\Line(30,70)(20,70)
\Line(30,70)(30,80)

\Line(70,70)(70,80)
\Line(70,70)(80,70)

\Text(-60,50)[l]{\bf SINGLET:}

\Text(13,15)[l]{$3^-$}
\Text(78,15)[l]{$4^+$}
\Text(7,72)[l]{$5^+$}
\Text(26,88)[l]{$6^+$}

\Text(66,88)[l]{$1^-$}
\Text(83,72)[l]{$2^-$}

\Text(35,40)[c]{$^+$}
\Text(35,55)[c]{$^-$}
\Text(42,64)[c]{$^-$}
\Text(60,64)[c]{$^+$}
\Text(66,55)[c]{$^+$}
\Text(66,40)[c]{$^-$}
\Text(60,33)[c]{$^+$}
\Text(42,33)[c]{$^-$}

\SetWidth{1.5}
\DashLine(50,20)(50,80){3}

\end{picture}\hspace{2cm}

\label{SingletNonsingletFigure}
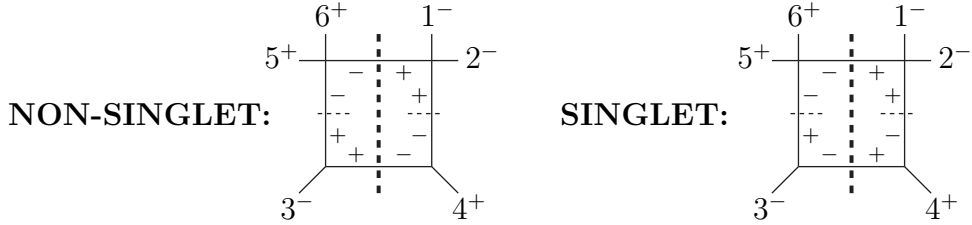
\caption{The non-singlet and singlet contributions to
a two-mass hard box integral. The dashed line indicates the cut to which the
singlet and non-singlet description refer.}
}

First consider the amplitude, $A_6(1^-, 2^-, 3^-, 4^+, 5^+, 6^+)$.
In this case, there will be a total of twelve non-vanishing integral
coefficients.
In this case many of the coefficients are equal to
each other,
$$
\eqalign{
c_{}^{(1^-2^-3^-)4^+5^+6^+} = c_{}^{(4^+5^+6^+)1^-2^-3^-}
= c_{}^{(2^-3^-)(4^+5^+)6^+1^-}
= c_{}^{(5^+6^+)(1^-2^-)3^-4^+} &= B_1\,, \nonumber \cr
c_{}^{(2^-3^-4^+)5^+6^+1^-} = c_{}^{(5^+6^+1^-)2^-3^-4^+}
= c_{}^{(3^-4^+)(5^+6^+)1^-2^-}
= c_{}^{(6^+1^-)(2^-3^-)4^+5^+} &= B_2\,, \nonumber\cr
c_{}^{(3^-4^+5^+)6^+1^-2^-} = c_{}^{(6^+1^-2^-)3^-4^+5^+}
= c_{}^{(4^+5^+)(6^+1^-)2^-3^-}
= c_{}^{(1^-2^-)(3^-4^+)5^+6^+} &= B_3\,,
\cr}
\equn\label{mmmppp}
$$
where the plus and minus labels on the legs refer to the helicity
labels in an all outgoing 
convention. For $B_1$ the non-singlet contribution  
vanishes whilst for the other two they are a sum on 
non-singlet and singlet contributions. 
From ref.~\cite{BDDKb}, the
explicit values are,
$$
\eqalign{
\null\hskip 0.5 truecm B_1 &= i \, { (K^2)^3
  \over \spb1.2\spb2.3\spa4.5\spa5.6\
\BR14 \BR36 }\,, \hskip 2.8 cm  K=K_{123}\,,  \cr
B_2 &= \left({ \BR41
         \over K_{}^2 } \right)^4  \ B_1 \vert_{j\to j+1}
       + \left({ \spa2.3\spb5.6 \over K^2_{} } \right)^4
            \ B_1^\cc \vert_{j\to j+1}\,, \hskip 1 cm  K=K_{234}\,,\cr
B_3 &= \left({ \BR63
        \over K^2_{} } \right)^4  \ B_1 \vert_{j\to j-1}
       + \left({ \spa1.2\spb4.5 \over K^2_{} } \right)^4
            \ B_1^\cc \vert_{j\to j-1}\,, \hskip 1 cm  K=K_{345}\,.
\cr}
\equn\label{Bdefanswer}
$$
where the gluon polarisation tensors have been
expressed in a spinor helicity~\cite{SpinorHelicity,TreeReview} basis.
In this formalism amplitudes are expressed in terms of spinor
inner-products,
$$
\spa{j}.{l} = \langle j^- | l^+ \rangle = \bar{u}_-(k_j) u_+(k_l)\,,
\hskip 2 cm
\spb{j}.{l} = \langle j^+ | l^- \rangle = \bar{u}_+(k_j) u_-(k_l)\, ,
\equn
\label{spinorproddef}
$$%
where $u_\pm(k)$ is a massless Weyl spinor with momentum $k$ and plus
or minus chirality.  With the normalisation used here,
$\spb{i}.{j} = \sign(k_i^0 k_j^0)\spa{j}.{i}^*$
so that,
$$
\spa{i}.{j} \spb{j}.{i} = 2 k_i \cdot k_j = s_{ij}\,.
\equn
$$
(Note that $\spb{i}.{j}$ defined in this way
differs by an overall sign from the notation commonly used in
twistor-space studies~\cite{WittenTopologicalString}.)

For $ A_{6;1}^{\NeqFour}(1^-,2^-,3^+,4^-,5^+,6^+)$, the box coefficients
are,
$$
\eqalign{
c_{}^{(1^-2^-3^+)4^-5^+6^+} = c_{}^{(4^-5^+6^+)1^-2^-3^+}
= c_{}^{(2^-3^+)(4^-5^+)6^+1^-}
= c_{}^{(5^+6^+)(1^-2^-)3^+4^-} &= D_1\,, \cr
c_{}^{(2^-3^+4^-)5^+6^+1^-} = c_{}^{(5^+6^+1^-)2^-3^+4^-}
= c_{}^{(3^+4^-)(5^+6^+)1^-2^-}
= c_{}^{(6^+1^-)(2^-3^+)4^-5^+} &= D_2\,, \cr
c_{}^{(3^+4^-5^+)6^+1^-2^-} = c_{}^{(6^+1^-2^-)3^+4^-5^+}
= c_{}^{(4^-5^+)(6^+1^-)2^-3^+}
= c_{}^{(1^-2^-)(3^+4^-)5^+6^+} &= D_3\,,
\cr}\equn
$$
where
$$
\eqalign{
 \null\hskip 0.5 truecm D_1 =& \left({
\BR34          \over K^2_{} } \right)^4    \ B_1
       + \left({ \spa1.2\spb5.6 \over K^2 } \right)^4
            \ B_1^\cc \ , \hskip 2.9cm   K=K_{123} \,, \cr
  D_2 =& \left({
\BR31         \over K^2 } \right)^4    \ B_1 \vert_{j\to j+1}
       + \left({ \spa2.4\spb5.6 \over K^2 } \right)^4
            \ B_1^\cc \vert_{j\to j+1}\ , \hskip .7cm  K=K_{234}\,, \cr
  D_3 =& \left({  \BR64
         \over K^2 } \right)^4    \ B_1 \vert_{j\to j-1}
       + \left({ \spa1.2\spb3.5 \over K^2_{345} } \right)^4
            \ B_1^\cc \vert_{j\to j-1}\,, \hskip .8 cm  K=K_{345}\,.
\cr}
\equn\label{DEFdef}
$$
Finally, for $A_{6;1}^{\NeqFour}(1^-,2^+,3^-,4^+,5^-,6^+)$,
$$\eqalign{
c_{}^{(1^-2^+3^-)4^+5^-6^+} = c_{}^{(4^+5^-6^+)1^-2^+3^-}
= c_{}^{(2^+3^-)(4^+5^-)6^+1^-}
= c_{}^{(5^-6^+)(1^-2^+)3^-4^+} &= G_1\,, \nonumber \cr
c_{}^{(2^+3^-4^+)5^-6^+1^-} = c_{}^{(5^-6^+1^-)2^+3^-4^+}
= c_{}^{(3^-4^+)(5^-6^+)1^-2^+}
= c_{}^{(6^+1^-)(2^+3^-)4^+5^-} &= G_2\,, \nonumber\cr
c_{}^{(3^-4^+5^-)6^+1^-2^+} = c_{}^{(6^+1^-2^+)3^-4^+5^-}
= c_{}^{(4^+5^-)(6^+1^-)2^+3^-}
= c_{}^{(1^-2^+)(3^-4^+)5^-6^+} &= G_3\,,
\cr}
\equn
$$
where
$$
\eqalign{
\null\hskip 0.5 truecm  G_1 &=  \left(
  {\BR25
         \over K^2 } \right)^4 \ B_1
       + \left({ \spa1.3\spb4.6 \over K^2 } \right)^4
            \ B_1^\cc  \,,\hskip 2.8 cm  K=K_{123} \, , \nonumber \cr
  G_2 &=\left( {\BR63
 \over K^2 } \right)^4    \ B_1^\cc \vert_{j\to j+1}
       + \left({ \spa5.1\spb2.4 \over K^2 } \right)^4
      \ B_1 \vert_{j\to j+1} \, , \hskip .6 cm  K=K_{234} \,,  \nonumber \cr
  G_3 &= \left( {\BR41
         \over K^2 } \right)^4    \ B_1^\cc \vert_{j\to j-1}
       + \left({ \spa3.5\spb6.2 \over K^2 } \right)^4
            \ B_1 \vert_{j\to j-1}\,, \hskip .3 cm \;\;\;  K=K_{345}  \,.
\cr}
\equn\label{GHKdef}
$$

Recently, the complete expression for all NMHV amplitudes in
$\NeqFour$ super-Yang-Mills theory have been obtained using the
unitarity method~\cite{BDKn}.  We will use the results of
that paper to construct some examples of $n$-point box coefficients
in the $\NeqEight$ theory with $n > 6$.
\
\section{Structure of One-Loop Gravity Amplitudes}
\label{GravityStructureSection}

For simplicity, in the forthcoming equations we
will define one-loop amplitudes in gravity for which all field
couplings have been removed, {\it i.e.},
$$
{\cal M}_n^{\rm 1-loop}
= \left( {\kappa \over 2} \right)^{n} M^{\rm 1-loop}_n(1,2,\ldots, n )
\;\; .
\equn
$$
In gravity theories, the three graviton vertex contains two powers of
momenta. A generic $n$-point diagram will involve a loop momenta
integral where the polynomial on the numerator is, in general,
of degree $2n$ in loop momenta. In $\NeqEight$ supergravity,
there are cancellations between the contributions of different
particle types, and in a suitable formulation the polynomial is only
of degree $2n-8$ in the loop momenta. (See, for example, the
``String Based Method''~\cite{StringBased,DunNor} where
the cancellation is explicit in terms of Feynman parameters.)
Recall, amplitudes are cut-constructible if the loop momenta
polynomial is of degree $\leq n-4$ and thus we would not expect
$\NeqEight$ supergravity amplitudes to be cut constructible
for $n > 6$.
The Passarino-Veltman~\cite{PassVelt} reduction, decreases a polynomial of
degree $r$ in a $n$-point integral to a box integral with polynomial
degree $r-(n-4)$. Hence for $n> 4$ we would {\it a priori} expect
tensor box integrals reduced to scalar boxes
plus triangle and bubble integrals.

However, as demonstrated in ref.~\cite{Bern:1998sv}, the one-loop MHV
amplitudes of $\NeqEight$ supergravity do have a much better power
behaviour than expected and appear to satisfy the 
cut constructibility criterion.  This was
demonstrated through six points by direct calculation and argued to
hold for all $n$ based on the factorisation properties.  Even more
surprisingly the $\NeqEight$ supergravity amplitudes appear to obey
precisely the same power counting as those of $\NeqFour$
super-Yang-Mills theory, {\it i.e.,} for an $n$-point integral there
are $n-4$ powers of loop momentum. This power counting behaviour is
reflected in the lack of triangle, bubble or additional rational
function contributions to MHV $\NeqEight$ supergravity amplitudes.

We do not have a proof that this feature holds more generally for
non-MHV amplitudes.  However, we conjecture that it does hold
generally for $\NeqEight$ one-loop amplitudes, using arguments
involving the factorisation properties of the amplitudes. Examining
their various factorisations, we find no evidence requiring integral
functions other than boxes to be present in the one-loop amplitudes.
In particular since the NMHV amplitudes can be reduced to MHV
amplitudes in various factorisation limits either the non-box
functions must all vanish in all such limits or, more likely, be
absent in NMHV amplitudes.

More explicitly, consider the multi-particle factorisations.  From
general field theory considerations, amplitudes must factorise (up to
subtleties having to do with infrared singularities) on multi-particle
poles.  For $K^\mu \equiv k_i^\mu+\ldots +k_{i+r+1}^\mu$ the
amplitude factorises when $K$ becomes
on shell.
Specifically, as $K^2 \rightarrow 0$ the factorisation
properties for one-loop infrared singular amplitudes are described
by~\cite{BernChalmers},
$$
\eqalign{
M_{n;1}^{\oneloop}\
& \hskip -.12 cm
 \mathop{\longrightarrow}^{K^2 \rightarrow 0}
\hskip .15 cm
\sum_{\lambda=\pm}  \Biggl[
   M_{r+1;1}^{\oneloop}(k_i, \ldots, k_{i+r-1}, K^\lambda) \, {i \over K^2} \,
   M_{n-r+1}^{\tree}((-K)^{-\lambda}, k_{i+r}, \ldots, k_{i-1}) \cr
& \hskip-1.5cm \null
 + M_{r+1}^{\tree}(k_i, \ldots, k_{i+r-1}, K^\lambda) \, {i\over K^2} \,
   M_{n-r+1;1}^{\oneloop}((-K)^{-\lambda}, k_{i+r}, \ldots, k_{i-1})
\label{LoopFact} \cr
& \hskip-1.5cm \null
 + M_{r+1}^{\tree}(k_i, \ldots, k_{i+r-1}, K^\lambda) \, {i\over K^2} \,
   M_{n-r+1}^{\tree}((-K)^{-\lambda}, k_{i+r}, \ldots, k_{i-1}) \,
      \cg\,  \Fact_n(K^2;k_1, \ldots, k_n) \Biggr] \,,
\cr}
\equn
$$
where the one-loop ``factorisation function'' $\Fact_n$ is
independent of helicities.   Consider the multi-particle
factorisation of a NMHV amplitude. In general, the two factorised
amplitudes appearing on either side of the pole will be MHV
amplitudes, since supersymmetry requires all amplitudes with less
than two negative helicities to vanish. From
ref.~\cite{Bern:1998sv} we know that the one-loop MHV amplitudes
should not contain triangle of bubble integrals.  Therefore, in {\it
all} multi-particle factorisation limits we cannot encounter these
integral functions. Moreover, in any soft or collinear limit which
reduces negative helicities by one, there can be no triangle or
bubble functions.  One can continue this bootstrap adding
increasing numbers of negative helicities. Although this does not
constitute a proof that triangle and bubble integrals cannot
appear we know of no counterexample, with six or higher points,
where a factorisation bootstrap has failed to produce the correct
result. In this case, the result is that bubble and triangle
integrals should not appear in one-loop $\NeqEight$ supergravity
amplitudes. In the remaining part of the paper we will assume that:

\begin{center}
{\it ``The only non-vanishing integral
functions in $\NeqEight$ one-loop amplitudes are scalar
 box integral functions'' }
\end{center}

\noindent
Thus, we write the $\NeqEight$ one-loop amplitudes
as
$$
M_{n}^{\NeqEight} = i \cg \, (\mu^2)^\e
\sum
c^{(i_1\ldots i_2) (i_3 \ldots i_4) (i_5 \ldots i_6) (i_7 \ldots i_8) }_{\NeqEight}
  F_{(i_1\ldots i_2) (i_3 \ldots i_4) (i_5 \ldots i_6) (i_7 \ldots i_8)}\,,
\equn\label{GenBoxDecomp}
$$
where the sum runs over all inequivalent box functions.  
The
$F_{(i_1\ldots i_2) (i_3\ldots i_4) (i_5 \ldots i_6) (i_7 \ldots i_8)}$
are dimensionally regulated box functions: although supergravity
theories are one-loop ultraviolet finite in four dimensions the
amplitudes contain infrared infinities requiring regularisation.
The
$c_{\NeqEight}^{(i_1\ldots i_2) (i_3 \ldots i_4) (i_5 \ldots i_6) (i_7
\ldots i_8) }$
are the kinematic coefficients, where the
parenthesis indicate which legs belong in a cluster as illustrated in
\fig{GravityCoefficientFigure}.  In the gravity case,
in contrast to the gauge theory case, the legs may 
appear in any ordering, since there is no colour ordering.

\FIGURE{
\begin{picture}(100,100)(0,0)

\Line(30,30)(30,70)
\Line(70,30)(70,70)
\Line(30,30)(70,30)
\Line(70,70)(30,70)

\Line(30,70)(20,70)
\Line(30,70)(30,80)

\Line(70,30)(70,20)
\Line(70,30)(80,30)

\Line(70,70)(70,80)
\Line(70,70)(80,70)

\Line(30,20)(30,30)
\Line(20,30)(30,30)

\Text(30,10)[c]{$i_7$}
\Text(10,30)[c]{$i_8$}
\Text(25,25)[c]{$\bullet$}
\Text(70,10)[c]{$i_6$}
\Text(90,30)[c]{$i_5$}
\Text(70,90)[c]{$i_3$}
\Text(90,70)[c]{$i_4$}
\Text(30,90)[c]{$i_2$}
\Text(10,70)[c]{$i_1$}
\Text(75,25)[c]{$\bullet$}
\Text(75,75)[c]{$\bullet$}
\Text(25,75)[c]{$\bullet$}

\end{picture}
\label{GravityCoefficientFigure}
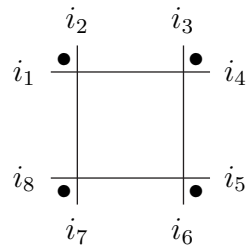
\caption{The labels corresponding to a box coefficient.  In the gravity
case there is no ordering imposed on the legs.}
}

\section{Constructing  Supergravity Amplitudes from
Super-Yang-Mills Amplitudes}
\label{GravityAmplitudesSection}

In this section, we obtain the box-coefficients of gravity amplitudes
using the KLT relations and the solutions for the Yang-Mills
amplitudes.  As examples, we will obtain all the box coefficients of
the six-graviton amplitude in $\NeqEight$ supergravity, as well as a
few selected coefficients at seven, eight and $n$ points.  The MHV
$n$-point amplitude have been previously been obtained in 
ref.~\cite{Bern:1998sv}, so here we focus on NMHV amplitudes. We also 
discuss one N$^2$MHV example at eight points.

\subsection{Six-Graviton NMHV Amplitudes}
\def\FIGdaveA{
\begin{picture}(100,100)(0,0)
\DashLine(50,73)(50,61){2}
\DashLine(50,39)(50,27){2}
\DashLine(27,50)(39,50){2}
\DashLine(61,50)(73,50){2}

\Line(30,30)(30,70)
\Line(70,30)(70,70)
\Line(30,30)(70,30)
\Line(70,70)(30,70)

\Line(30,30)(20,20)
\Line(70,30)(80,20)

\Line(30,70)(20,70)
\Line(30,70)(30,80)

\Line(70,70)(70,80)
\Line(70,70)(80,70)

\Text(13,15)[l]{$1^-$}
\Text(78,15)[l]{$6^+$}
\Text(7,72)[l]{$2^-$}
\Text(26,88)[l]{$3^-$}

\Text(66,88)[l]{$4^+$}
\Text(83,72)[l]{$5^+$}

\Text(35,40)[c]{$^-$}
\Text(35,55)[c]{$^+$}
\Text(42,64)[c]{$^+$}
\Text(60,64)[c]{$^-$}
\Text(66,55)[c]{$^-$}
\Text(66,40)[c]{$^+$}
\Text(60,33)[c]{$^-$}
\Text(42,33)[c]{$^+$}

\Text(20,52)[c]{$\ell_1$}
\Text(52,20)[c]{$\ell_6$}
\Text(80,52)[c]{$\ell_5$}
\Text(52,80)[c]{$\ell_3$}
\end{picture}
}

\def\FIGdaveB{
\begin{picture}(100,100)(0,0)
\DashLine(50,73)(50,61){2}
\DashLine(50,39)(50,27){2}
\DashLine(27,50)(39,50){2}
\DashLine(61,50)(73,50){2}

\Line(30,30)(30,70)
\Line(70,30)(70,70)
\Line(30,30)(70,30)
\Line(70,70)(30,70)

\Line(30,30)(20,20)
\Line(70,30)(80,20)

\Line(30,70)(20,70)
\Line(30,70)(30,80)

\Line(70,70)(70,80)
\Line(70,70)(80,70)

\Text(13,15)[l]{$3^-$}
\Text(78,15)[l]{$4^+$}
\Text(7,72)[l]{$5^+$}
\Text(26,88)[l]{$6^+$}

\Text(66,88)[l]{$1^-$}
\Text(83,72)[l]{$2^-$}

\Text(35,40)[c]{$^+$}
\Text(35,55)[c]{$^-$}
\Text(42,64)[c]{$^-$}
\Text(60,64)[c]{$^+$}
\Text(66,55)[c]{$^+$}
\Text(66,40)[c]{$^-$}
\Text(60,33)[c]{$^+$}
\Text(42,33)[c]{$^-$}

\Text(20,52)[c]{$\ell_3$}
\Text(52,20)[c]{$\ell_4$}
\Text(80,52)[c]{$\ell_2$}
\Text(52,80)[c]{$\ell_6$}
\end{picture}
\begin{picture}(100,100)(0,0)
\DashLine(50,73)(50,61){2}
\DashLine(50,39)(50,27){2}
\DashLine(27,50)(39,50){2}
\DashLine(61,50)(73,50){2}

\Line(30,30)(30,70)
\Line(70,30)(70,70)
\Line(30,30)(70,30)
\Line(70,70)(30,70)

\Line(30,30)(20,20)
\Line(70,30)(80,20)

\Line(30,70)(20,70)
\Line(30,70)(30,80)

\Line(70,70)(70,80)
\Line(70,70)(80,70)

\Text(13,15)[l]{$3^-$}
\Text(78,15)[l]{$4^+$}
\Text(7,72)[l]{$5^+$}
\Text(26,88)[l]{$6^+$}

\Text(66,88)[l]{$1^-$}
\Text(83,72)[l]{$2^-$}

\Text(35,40)[c]{$^+$}
\Text(35,55)[c]{$^-$}
\Text(42,64)[c]{$^-$}
\Text(60,64)[c]{$^+$}
\Text(66,55)[c]{$^+$}
\Text(66,40)[c]{$^-$}
\Text(60,33)[c]{$^-$}
\Text(42,33)[c]{$^+$}

\Text(20,52)[c]{$\ell_3$}
\Text(52,20)[c]{$\ell_4$}
\Text(80,52)[c]{$\ell_2$}
\Text(52,80)[c]{$\ell_6$}
\end{picture}
}

\def\FIGdaveC{
\begin{picture}(100,100)(0,0)
\DashLine(50,73)(50,61){2}
\DashLine(50,39)(50,27){2}
\DashLine(27,50)(39,50){2}
\DashLine(61,50)(73,50){2}

\Line(30,30)(30,70)
\Line(70,30)(70,70)
\Line(30,30)(70,30)
\Line(70,70)(30,70)

\Line(30,30)(20,20)
\Line(70,30)(80,20)

\Line(30,70)(20,70)
\Line(30,70)(30,80)

\Line(70,70)(70,80)
\Line(70,70)(80,70)

\Text(13,15)[l]{$6^+$}
\Text(78,15)[l]{$3^-$}
\Text(7,72)[l]{$1^-$}
\Text(26,88)[l]{$4^+$}

\Text(66,88)[l]{$2^-$}
\Text(83,72)[l]{$5^+$}

\Text(35,40)[c]{$^+$}
\Text(35,55)[c]{$^-$}
\Text(42,64)[c]{$^+$}
\Text(60,64)[c]{$^-$}
\Text(66,55)[c]{$^+$}
\Text(66,40)[c]{$^-$}
\Text(60,33)[c]{$^+$}
\Text(42,33)[c]{$^-$}

\Text(20,52)[c]{$\ell_6$}
\Text(52,20)[c]{$\ell_3$}
\Text(80,52)[c]{$\ell_5$}
\Text(52,80)[c]{$\ell_4$}
\end{picture}
\begin{picture}(100,100)(0,0)
\DashLine(50,73)(50,61){2}
\DashLine(50,39)(50,27){2}
\DashLine(27,50)(39,50){2}
\DashLine(61,50)(73,50){2}

\Line(30,30)(30,70)
\Line(70,30)(70,70)
\Line(30,30)(70,30)
\Line(70,70)(30,70)

\Line(30,30)(20,20)
\Line(70,30)(80,20)

\Line(30,70)(20,70)
\Line(30,70)(30,80)

\Line(70,70)(70,80)
\Line(70,70)(80,70)

\Text(13,15)[l]{$6^+$}
\Text(78,15)[l]{$3^-$}
\Text(7,72)[l]{$1^-$}
\Text(26,88)[l]{$4^+$}

\Text(66,88)[l]{$2^-$}
\Text(83,72)[l]{$5^+$}

\Text(35,40)[c]{$^-$}
\Text(35,55)[c]{$^+$}
\Text(42,64)[c]{$^-$}
\Text(60,64)[c]{$^+$}
\Text(66,55)[c]{$^-$}
\Text(66,40)[c]{$^+$}
\Text(60,33)[c]{$^-$}
\Text(42,33)[c]{$^+$}

\Text(20,52)[c]{$\ell_6$}
\Text(52,20)[c]{$\ell_3$}
\Text(80,52)[c]{$\ell_5$}
\Text(52,80)[c]{$\ell_4$}

\end{picture}
\begin{picture}(100,100)(0,0)
\DashLine(50,73)(50,61){2}
\DashLine(50,39)(50,27){2}
\DashLine(27,50)(39,50){2}
\DashLine(61,50)(73,50){2}

\Line(30,30)(30,70)
\Line(70,30)(70,70)
\Line(30,30)(70,30)
\Line(70,70)(30,70)

\Line(30,30)(20,20)
\Line(70,30)(80,20)

\Line(30,70)(20,70)
\Line(30,70)(30,80)

\Line(70,70)(70,80)
\Line(70,70)(80,70)

\Text(13,15)[l]{$6^+$}
\Text(78,15)[l]{$3^-$}
\Text(7,72)[l]{$1^-$}
\Text(26,88)[l]{$4^+$}

\Text(66,88)[l]{$2^-$}
\Text(83,72)[l]{$5^+$}

\Text(35,40)[c]{$^-$}
\Text(35,55)[c]{$^+$}
\Text(42,64)[c]{$^-$}
\Text(60,64)[c]{$^+$}
\Text(66,55)[c]{$^-$}
\Text(66,40)[c]{$^+$}
\Text(60,33)[c]{$^+$}
\Text(42,33)[c]{$^-$}

\Text(20,52)[c]{$\ell_6$}
\Text(52,20)[c]{$\ell_3$}
\Text(80,52)[c]{$\ell_5$}
\Text(52,80)[c]{$\ell_4$}

\end{picture}
}

\def\FigDaveD{
\begin{picture}(100,100)(0,0)
\DashLine(50,73)(50,61){2}
\DashLine(50,39)(50,27){2}
\DashLine(27,50)(39,50){2}
\DashLine(61,50)(73,50){2}

\Line(30,30)(30,70)
\Line(70,30)(70,70)
\Line(30,30)(70,30)
\Line(70,70)(30,70)
\Line(30,30)(20,30)
\Line(30,30)(30,20)
\Line(70,30)(80,20)
\Line(70,70)(80,80)
\Line(30,70)(20,80)
\Text(10,90)[l]{$1^-$}
\Text(83,90)[l]{$2^+$}
\Text(83,15)[l]{$3^+$}
\Text(25,15)[l]{$4^+$}
\Text(10,33)[l]{$5^-$}
\Text(35,40)[c]{$^+$}
\Text(42,33)[c]{$^+$}
\Text(35,55)[c]{$^-$}
\Text(42,64)[c]{$^+$}
\Text(66,55)[c]{$^-$}
\Text(60,64)[c]{$^-$}
\Text(66,40)[c]{$^+$}
\Text(60,33)[c]{$^-$}
\Text(20,52)[c]{$\ell_5$}
\Text(52,20)[c]{$\ell_3$}
\Text(80,52)[c]{$\ell_2$}
\Text(52,80)[c]{$\ell_1$}
\end{picture}
}

In this subsection, we obtain the box-coefficients of the NMHV
$\NeqEight$ six-graviton amplitude in terms of the, known, box
coefficients of $\NeqFour$ super-Yang-Mills, collected in
\sec{6ptYMSubsection}.  There are three independent NMHV amplitudes in
Yang-Mills depending on the positions of the three negative legs:
$A(1^-2^-3^-4^+5^+6^+)$, $A(1^-2^-3^+4^-5^+6^+)$ and
$A(1^-2^+3^-4^+5^-6^+)$.  In gravity, on the other hand, there is no
colour ordering and hence a single distinct amplitude:
$M(1^-2^-3^-4^+5^+6^+)$.  The box-coefficients of the gravity
amplitude will be expressible in terms of the box coefficients of the
three Yang-Mills amplitudes.  For six-point amplitudes, there are
three possible scalar box structures in the amplitude: the
``single-mass'', ``two-mass-hard'' and ``two-mass-easy'' boxes.  (See,
{\it e.g.,} the first appendix of ref.~\cite{BDDKb}, for definitions of
these functions in the Euclidean region.)  Of these only the first two
contribute to the super-Yang-Mills six-gluon NMHV amplitude. This property is
immediately ``inherited'' by the gravity case. If we evaluate the
appropriate generalised cut and express the gravity amplitudes in
terms of Yang-Mills we immediately find that the gravity coefficient
vanishes whenever the corresponding super-Yang-Mills coefficient
vanishes. For the ``two-mass-hard'' coefficients there
are total of 45 independent boxes all of which appear in the
amplitude. These can be split into four distinct cases depending on
the position of the three negative legs,
$$
I^{(--),(++),+-}\,,\,\,
I^{(--),(++),-+}\,,\,\,
I^{(--),(-+),++}\,,\,\,
I^{(-+),(-+),-+}\, .
\equn
$$
All other coefficients of two-mass-hard boxes are obtained by symmetry
and parity conjugation.

Taking the first case it is straightforward to show
that only a single helicity configuration of the internal
lines contributes to the quadruple cut
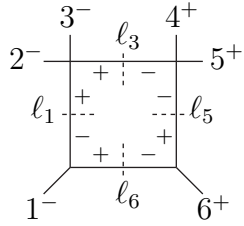
\begin{center}
\FIGdaveA
\end{center}
\noindent
In this case, gravitons are the only 
possible states when all cut conditions
are imposed. The coefficient of the box function is thus
$$
\eqalign{
\hat c_{\NeqEight}^{(2^-3^-)(4^+5^+)6^+1^-}
& =
\half
M^\tree(2^-,3^-,-\ell_1^+,\ell_3^+) \times
M^\tree(4^+,5^+,-\ell_3^-,\ell_5^-)\cr & \hspace{3cm}\times
M^\tree(6^+,-\ell_5^+,\ell_6^-) \times
M^\tree(1^-,-\ell_6^+,\ell_1^-)
\cr
& = \half
 s_{23} A^\tree(2^-,3^-,-\ell_1^+,\ell_3^+) A^\tree(3^-,2^-,-\ell_1^+,\ell_3^+)
\cr 
& \hspace{3.cm}\times s_{45} A^\tree(4^+,5^+,-\ell_3^-,\ell_5^-)
A^\tree(5^+,4^+,-\ell_3^-,\ell_5^-)
\cr & 
\hskip .4 cm 
\times A^\tree(6^+,-\ell_5^+,\ell_6^-)A^\tree(6^+,-\ell_5^+,\ell_6^-) \cr
& \hskip 3 cm 
\times A^\tree(1^-,-\ell_6^+,\ell_1^-)A^\tree(1^-,-\ell_6^+,\ell_1^-)
\cr &
= \half s_{23}s_{45}  \times
\biggl( A^\tree(2^-,3^-,-\ell_1^+,\ell_3^+) \times
A^\tree(4^+,5^+,-\ell_3^-,\ell_5^-) \cr & \hspace{3cm} \times
A^\tree(6^+,-\ell_5^+,\ell_6^-) \times
A^\tree(1^-,-\ell_6^+,\ell_1^-) \biggr)
\cr & \hskip .4 cm 
\times
\biggl( A^\tree(3^-,2^-,-\ell_1^+,\ell_3^+) \times
A^\tree(5^+,4^+,-\ell_3^-,\ell_5^-) \times \cr & \hspace{3cm}
A^\tree(6^+,-\ell_5^+,\ell_6^-) \times
A^\tree(1^-,-\ell_6^+,\ell_1^-) \biggr)
\cr
& =  2 s_{23}s_{45} \times \hat c_S^{(2^-3^-)(4^+5^+)6^+1^-}
 \times  \hat c_S^{(3^-2^-)(5^+4^+)6^+1^-}\,,
\cr} \equn
$$
where the $\NeqFour$ box integral coefficients $\hat c$ are defined in
subsection~\ref{6ptYMSubsection}.  In going to the last line, we used
the fact that the quadruple cut freezes the loop
integral~\cite{BrittoUnitarity}, determining the $\NeqFour$
coefficients directly.  After substituting for the explicit form of
the $\hat c$, using \eqns{hatcvsc}{mmmppp} and relabeling,  we obtain,
$$
\eqalign{
 & \hat c_{\NeqEight}^{(2^-3^-)(4^+5^+)6^+1^-}
=
\cr
& \hspace{0.3cm}
{ i\over 2} { s_{23}s_{45} s_{61}^2 (K_{123}^2)^8 \over
 \spb1.2\spb2.3\spb1.3\spb3.2
\spa4.5\spa5.6\spa5.4\spa4.6
\BR14\BR15
\BR26\BR36    }\,.
\cr}\equn
 $$

Consider now the coefficient $\hat
c_{\NeqEight}^{(5^+6^+)(1^-2^-)4^+3^-}$.  Here there are two possible
helicity configurations in the generalized cuts,
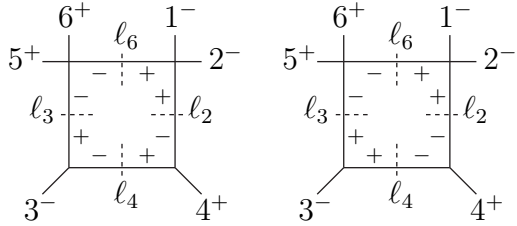
\begin{center}
\FIGdaveB
\end{center}
where, again, only gravitons may contribute. For the Yang-Mills
case the two configurations give rise to the non-singlet and
singlet terms in the box
coefficients. For the gravity case we take each configuration separately and
decompose the gravity amplitudes into Yang-Mills amplitudes. 
The gravity coefficient will then be
$$
\hat c_{\NeqEight}^{(5^+6^+)(1^-2^-)4^+3^-}
 =  2  s_{56}s_{12} \biggl(
\sum_{i=NS,S} \hat c_{i}^{(5^+6^+)(1^-2^-)4^+3^-}
 \times \hat c_{i}^{(6^+5^+)(2^-1^-)4^+3^-}
\biggr)\,,
\equn
$$
Explicitly, this evaluates to
$$
\eqalign{
 {i\over 2} { \BR43^8s_{12}s_{56}s_{34}^2  \over
 \spa3.5 \spa3.6 \spa5.6 \spa6.5 \spb1.2 \spb1.4 \spb2.1
\spb2.4 \BR13\BR23\BR45\BR46}
\cr
\null +
{i\over 2}{ \spa1.2^6 \spb5.6^6s_{12}s_{56}s_{34}^2
\over  \spa1.4 \spa2.4 \spb3.5 \spb3.6 \BR31\BR32\BR54\BR64} ;
\;\;\;\;   K=K_{124}\,,
\cr}\equn\label{SixPointCaseB}
$$
Similarly we have
$$\hspace{1cm}
\eqalign{
\hat c_{\NeqEight}^{(2^-3^-)(4^+1^-)5^+6^+}
& =  2  s_{23}s_{41} \times
\biggl( \sum_{i=NS,S}
\hat c_i^{(2^-3^-)(4^+1^-)5^+6^+}
 \times \hat c_i^{(3^-2^-)(1^-4^+)5^+6^+}
\biggr)\,,
\cr
\hat c_{\NeqEight}^{(3^-4^+)(5^+6^+)1^-2^-}
& =   2 s_{34}s_{56}
\times
\biggl(
\sum_{i=NS,S}
\hat c_i^{(3^-4^+)(5^+6^+)1^-2^-}
 \times \hat c_i^{(4^+3^-)(6^+5^+)1^-2^-}
\biggr)\,.
\cr}\equn
$$

The final two-mass box case is
$\hat c_{\NeqEight}^{(1^-4^+)(2^-5^+)3^-6^+}$. In this case there are three
possible solutions,
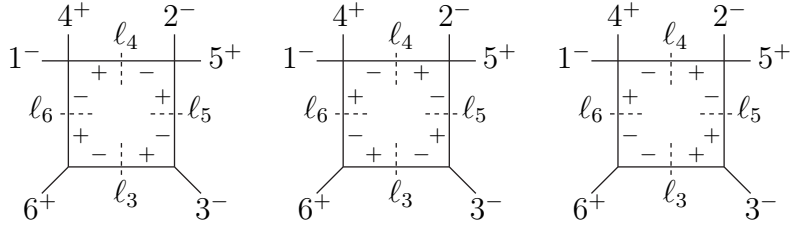
\begin{center}
\FIGdaveC
\end{center}

In this example, we must also consider the contributions from all the
states of the $\NeqEight$ multiplet to the first and second
diagrams. The combination of these different particle contributions
follows closely from the Yang-Mills case. For the $\NeqFour$
super-Yang-Mills case the MHV tree amplitudes for states of different
helicity can be related to the tree amplitude with two gluons and two
scalars by, for example,
$$
A^\tree(1^+,2^-,\ell_1^h,\ell_2^{-h}) =
 x^{2h} A^\tree(1^+,2^-,\ell_1^s,\ell_2^{-s})\,,
\equn
$$
by application of supersymmetric Ward identities~\cite{SWI}. By
applying the supersymmetric Ward identities at each corner the
contributions to the quadruple cut from each particle type will be
related to that of the scalar by an overall factor of $X^{2h}$
where
$$
X= x_1x_2x_3x_4\,.
\equn
$$
Summing over all the possible helicities over the first and second 
configurations gives a total coefficient,
which is that of the scalar multiplied by
$$
\rho_{\NeqFour} = \sum_{h\in H} X^{2h} = { (X-1)^4 \over X^2 }\,.
\equn
$$
When we carry out the same procedure for $\NeqEight$ supergravity the
argument follows in a very similar way (see ref.~\cite{BDDPR} for 
the details of summing over the $\NeqEight$ multiplet) 
with the contributions from the entire multiplet, summed over the two
configurations, being that of the scalar times
$$
\rho_{\NeqEight} = (\rho_{\NeqFour})^2\, . 
\equn
$$ This gives this contribution to the box coefficient as the product
of the non-singlet Yang-Mills terms. The third solution gives a
contribution which is the product of the Yang-Mills singlet terms so
that, adding the contributions together gives
$$\hspace{1.3cm}
\hat c_{\NeqEight}^{(1^-4^+)(2^-5^+)3^-6^+}
= 2 s_{14}s_{25} \times \biggl( \sum_{i=NS,S}
\hat c_{i}^{(1^-4^+)(2^-5^+)3^-6^+}
\times
\hat c_{i}^{(4^+1^-)(5^+2^-)3^-6^+} \biggr)\,.
\equn
$$

In general, if we consider boxes with massive legs containing more
than two external legs, then the KLT relationships will express the
gravity tree as a sum of products of Yang-Mills trees. Inserting
this into the generalised cuts the $\NeqEight$ box coefficient will be a
sum over products of $\NeqFour$ coefficients
$$
\hat c_{\NeqEight} =   \sum P(s_{ij})\hat c_{\NeqFour}
 \, \hat c'{}_{\NeqFour}\,,
\equn
$$
where $P(s_{ij})$ is a polynomial in the $s_{ij}$. 

Continuing with the six-point amplitude we now consider the
coefficients of the ``one-mass'' box.  There are a total of 60
such box functions. These split into four classes depending on
whether the massive leg contains three, two, one or no negative
helicities. Considering the case of the massive leg containing three
negative helicities, we have
$$\eqalign{
\hat c^{(1^-2^-3^-)4^+5^+6^+}_{\NeqEight} 
&=\half M^\tree(1^-, 2^-, 3^-, -\ell_6^+, \ell_3^+)
\times M^\tree( 4^+ , -\ell_3^-, \ell_4^{+} )\cr& 
\hspace{4cm} \times M^\tree( 5^+ , -\ell_4^{-},  \ell_5^{-}
) \times M^\tree( 6^+ , -\ell_5^{+}, \ell_6^- )\,.
\cr}\equn
$$
Now the KLT expansion
for $M^\tree(1^-, 2^-, 3^-, -\ell_6^+, \ell_3^+)$ gives
$$
\eqalign{
M^\tree(1^-, 2^-, 3^-, \ell_3^+, -\ell_6^+)
& =
i s_{12}s_{3\ell_3} A^\tree(1^-, 2^-, 3^-, \ell_3^+, -\ell_6^+)
A^\tree(2^-, 1^-, \ell_3^+, 3^-, -\ell_6^+)
\cr
& \null \hskip .1 truecm +i s_{13}s_{2\ell_3} A^\tree(1^-, 3^-, 2^-, \ell_3^+, -\ell_6^+)
 A^\tree(3^-, 1^-, \ell_3^+, 2^-, -\ell_6^+)\,.
\cr} \equn
$$
This expression contains Yang-Mills amplitudes where the cut legs
$\ell_6$ and $\ell_3$ are not adjacent.  To recombine these into
Yang-Mills box coefficients we must remedy this by using the
``decoupling identity'' among colour-ordered tree
amplitudes~\cite{Colour}
$$
\Atree(a,\{ \alpha \},b,\{ \beta\} ) =(-1)^{n_\beta} \sum_{
\sigma\in OP\{ \alpha \}\{ \beta^T \}} \Atree(a, \sigma( \{ \alpha
\} \{ \beta^T \}), b)\,,
\equn$$
where $n_\beta$ is the number of
elements in set $\{\beta\}$, $\beta^T$ is the set $\beta$ with
ordering reversed, and $OP\{ \alpha \}\{ \beta^T \}$ is the set of
permutation of $\{ \alpha \}\cup \{ \beta^T \}$ preserving the
ordering of elements within each of the two sets. The decoupling
identity, with $a= \ell_3, b=-\ell_6, \{\alpha\} = \{1\}$ and 
$\{\beta\} = \{2,1\}$,  implies 
$$
\eqalign{
\Atree(2^-, 1^-, \ell_3^+, 3^-, -\ell_6^+)
&=\Atree(\ell_3^+, 3^-, -\ell_6^+,2^-, 1^-)
\cr
&=\Big(\Atree(\ell_3^+,3^-,1^-,2^-, -\ell_6^+)
\cr & \hspace{0.6cm}+\Atree(\ell_3^+,1^-,3^-,2^-, -\ell_6^+)
+\Atree(\ell_3^+,1^-,2^-,3^-, -\ell_6^+)\Big)\,,
\cr
\Atree(3^-, 1^-, \ell_3^+, 2^-, -\ell_6^+)
&=\Atree(\ell_3^+, 2^-, -\ell_6^+,3^-, 1^-)
\cr
&=\Big(\Atree(\ell_3^+,2^-,1^-,3^-, -\ell_6^+)
\cr & \hspace{0.6cm}+\Atree(\ell_3^+,1^-,2^-,3^-, -\ell_6^+)
+\Atree(\ell_3^+,1^-,3^-,2^-, -\ell_6^+)\Big)\\.
\cr}
\equn$$
Recombining the products of Yang-Mills tree amplitudes 
into Yang-Mills box coefficients, we obtain for the gravity box coefficient
$$\hspace{2cm}
\eqalign{
\hat c_{\NeqEight}^{(1^-2^-3^-)4^+5^+6^+} &
= 2 s_{12}s_{3\ell_3}
\biggl( \hat c_{S}^{(1^-2^-3^-)4^+5^+6^+}
\hat c_{S}^{(2^-1^-3^-)4^+5^+6^+}
\cr
&\null\hspace{1.7cm}
+  \hat c_{S}^{(1^-2^-3^-)4^+5^+6^+}
\hat c_{S}^{(2^-3^-1^-)4^+5^+6^+}
\cr
&\null\hspace{1.7cm}
+
\hat c_{S}^{(1^-2^-3^-)4^+5^+6^+}
\hat c_{S}^{(3^-2^-1^-)4^+5^+6^+}
\biggr)
\cr
&
\hspace{0.3cm}+ 2 s_{13}s_{2\ell_3}
\biggl( \hat c_{S}^{(1^-3^-2^-)4^+5^+6^+}
\hat c_{S}^{(3^-1^-2^-)4^+5^+6^+} \cr
&\null\hspace{1.7cm}  +
\hat c_{S}^{(1^-3^-2^-)4^+5^+6^+}
\hat c_{S}^{(3^-2^-1^-)4^+5^+6^+}
\cr
&\null\hspace{1.7cm}  +
 \hat c_{S}^{(1^-3^-2^-)4^+5^+6^+}
\hat c_{S}^{(2^-3^-1^-)4^+5^+6^+}
\biggr)\,,
\cr}
\equn$$
where we replace
$$
s_{a\ell_3}
={ s_{56} \la 5^+ | a | 4^+ \ra \over \spb5.6\spa6.4 }
+s_{a4}\,,
\equn
$$
which is 
obtained by applying the on-shell conditions to the generalised
cuts~\cite{BrittoUnitarity}. (If the massless corner attached to leg four ($4$) was a ``mostly-minus'' three point amplitude
rather than a ``mostly-plus'' amplitudes then solving the on-shell conditions gives a formula for 
$s_{a\ell_3}$ which is the complex conjugate of the above.) 
The other one-mass box coefficients obey
an analogous formul\ae\ involving a summation over the singlet and
non-singlet solutions.

Summarising, we have shown that the $\NeqEight$ coefficients are given, in terms of the 
$\NeqFour$ super-Yang-Mills coefficients, by
$$
\eqalign{
\hat c_{N=8}^{(ab)c(de)f} &=  0\,,
\cr
\hat c_{N=8}^{(ab)(cd)ef} & = 
 2 s_{ab}s_{cd} \times \biggl( \sum_{i=NS,S}
\hat c_{i}^{(ab)(cd)ef}
\times
\hat c_{i}^{(ba)(dc)ef} \biggr)\,,
\cr
\hat c_{N=8}^{(abc)def} & =\null  2 s_{ab}s_{c\ell_c} \sum_{i=NS,S}
\biggl(\hat c_{i}^{(abc)def}
\hat c_{i}^{(bac)def}
+\hat c_{i}^{(abc)def}\hat c_{i}^{(bca)def}
+\hat c_{i}^{(abc)def} \hat c_{i}^{(cba)def}
\biggr)
\cr
&
\; \; + 2 s_{ac}s_{b\ell_c} \sum_{i=NS,S}
\biggl( \hat c_{i}^{(acb)def} \hat c_{i}^{(cab)def}
+\hat c_{i}^{(acb)def}\hat c_{i}^{(cba)def}
+\hat c_{i}^{(acb)def}\hat c_{i}^{(bca)def}
\biggr)\, ,
\cr}
\equn
$$
for all choices of helicities.  
Expressing the box coefficients in terms of $N=4$ super-Yang-Mills box-coefficients is useful
because it allows one to exploit the recent progress in computing such coefficients however it may not always give the
most compact realisation of the supergravity box coefficients. 
If our ansatz that only box integrals
contribute is correct, these coefficients give the complete NMHV
$\NeqEight$ one-loop six-point amplitudes.

\subsection{Sample Seven- and Eight-Point Box Coefficients}

We can use the generalised cuts together with the KLT relationship to generate
any supergravity or gravity amplitude from the equivalent
Yang-Mills amplitudes. For example consider one of the
three-mass boxes of the seven-point amplitude,
$$
\eqalign{
\hat c_{\NeqEight}^{(4^+5^+)(2^-3^-)(6^+7^+)1^-}
&=
2 s_{23}s_{45}s_{67} \hat c_{\NeqFour}^{(4^+5^+)(2^-3^-)(6^+7^+)1^-}
\hat c_{\NeqFour}^{(5^+4^+)(3^-2^-)(7^+6^+)1^-}
\cr
&
\null \hskip -3.4 truecm
=  {i \over 2}  {s_{45}s_{23}s_{67}
\Bigr( \spa2.3  \la  1^- | K_{67}K_{45}| 1^+ \ra \Bigl)^8
\spa5.2\spa3.6 \spa4.3\spa2.7  (  K_{671}^2K_{145}^2 -s_{45}s_{67})^2
\over 
N(1452367)N(1543276)  s_{23}^2
\prod_{j=2,3,6,7}  \la  1^- | K_{45}K_{23}| j^+ \ra
\prod_{j=2,3,4,5}  \la  1^- | K_{67}K_{23}| j^+ \ra
}\,,
\cr}
\equn\label{SevenPointExample}
$$
with $N(abcdemn)=\spa{a}.{b}\spa{b}.{c}\ldots \spa{n}.{a}$.  
For this configuration the $\NeqFour$ box-coefficients are entirely 
from singlet contributions. 
We used the results of refs.~\cite{BDDK7,BDKn} to substitute for the explicit
values of the $\NeqFour$ gauge theory coefficients.  We shall use this
result later when we explore the twistor structure of the NMHV
amplitudes.

A sample coefficient in  the eight-point amplitude is
$$
\hat c_{\NeqEight}^{(1^-2^-)(5^+6^+)(3^-4^-)(7^+8^+)}
=2 s_{12}s_{34}s_{45}s_{56}\, \hat
c_{\NeqFour}^{(1^-2^-)(5^+6^+)(3^-4^-)(7^+8^+)} \hat
c_{\NeqFour}^{(2^-1^-)(6^+5^+)(4^-3^-)(8^+7^+)}\,, \equn
$$ 
which gives us an example of an  N$^2$MHV box coefficient.  The
$\NeqFour$ coefficients in this formula may be obtained using the
results of ref.~\cite{BrittoUnitarity}.

Depending on how we decide to write out the gravity coefficients in
terms of Yang-Mills trees, we can have different forms of $\NeqEight$
box coefficients.  As a specific example of this is, an alternative
form of the coefficient is, {\it e.g.},
$$
\hat c_{\NeqEight}^{(1^-2^-)(5^+6^+)(3^-4^-)(7^+8^+)}
= 2 s_{12}s_{34}s_{45}s_{56}
\hat c_{\NeqFour}^{(1^-2^-)(6^+5^+)(3^-4^-)(8^+7^+)}
\hat c_{\NeqFour}^{(2^-1^-)(5^+6^+)(4^-3^-)(7^+8^+)}\,.
\equn
$$
It is interesting to note that the equivalence of the different forms  
implies a quadratic identity amongst the $\NeqFour$ box-coefficients.

\subsection{A Sample Gravity $n$-point  Coefficient}

In this section we present the computation of an $n$-point sample
term to illustrate the general process. We consider the specific
case of the one-mass box $I^{(4^+....n^+)1^-2^-3^-}$. 
The
quadruple cut in this case has the single, singlet, solution
$$
\eqalign{
\hat c^{(4^+,\ldots,n^+)1^-2^-3^-}_{\NeqEight} & = 
\half M^\tree(4^+,\ldots,n^+, \ell_n^-,   -\ell_3^-) \times
M^\tree(1^- , -\ell_n^+, \ell_1^{-} )  \cr
&\hspace{2.cm} 
\times M^\tree(2^- , -\ell_1^{+},  \ell_2^{+} )
\times M^\tree( 3^- , -\ell_2^{-}, \ell_3^+ )\, .
\cr}
\equn
$$
For definiteness, consider the case where $n=2m+3$ whence we can use
the following expression for the KLT relationships~\cite{Bern:1998sv},
$$
\eqalign{ 
M^\tree( -\ell_3^- ,4^+,  & \ldots,n^+, \ell_n^-) =  -i\biggl[
\Atree(-\ell_3^- ,4^+,\ldots,n^+, \ell_n^-)  \cr &
\null\hskip
-1.2truecm \times 
\sum_{ \alpha\in {\cal S}_{m},\beta\in {\cal S}_{m-1}} 
f(\alpha) \bar
f(\beta) \times
\Atree(\alpha_1,\ldots,\alpha_{m},-\ell_3,n,\beta_1,\ldots\beta_{m-1}, \ell_n )
\biggr]\cr 
& \hskip 2 cm \null 
 +{\cal P} (4,\ldots, n-1)\,. \cr}
\equn
$$
The $\alpha$ is a permutation of the $m$ legs 
$(4,\cdots
m+3)$ and $\beta$ is a permutation of the $m-1$ legs
$(m+4,\ldots n-1)$.
The functions $f(\alpha)$ and $\bar
f(\beta)$ are polynomial in momenta with
$$
\eqalign{
f(\alpha)
&= s(\ell_3,\alpha_m)\prod_{r=1}^{m-1}
\biggl( s(\ell_3,\alpha_r) +
\sum_{k=r+1}^{m}  g(\alpha_r,\alpha_k)  \biggr)\,,
\cr
\bar f(\beta)
& =
 s(\beta_1,n) \prod_{r=2}^{m-1}
\biggl( s(\beta_r,n)+ \sum_{k=1}^{r-1} g(\beta_k,\beta_r)
\biggr)\,,
\cr}
\equn$$
with
$$
 g(i,j)\ =\
\left(
 {   s(i,j) \equiv s_{ij}, \;\;  i>j, \atop
                                       0, \;\; {\rm otherwise.} }
\right)\,.
\equn
$$
Solving the on-shell conditions means we substitute
$$
\eqalign{
s(a, \ell_3) &={ s_{12} \la 4^+ | a | 2^+ \ra \over \spb4.1\spa1.2 }
-s_{a3}\, ,
\cr
s(a, \ell_n) &={ s_{23} \la 1^+ | a | 2^+ \ra \over \spb1.3\spa3.2 }
+s_{a1}\,.
\cr} \equn
$$

Before combining into $\NeqEight$ coefficients we must reorganise
$$
\eqalign{
\Atree(\alpha_1,\ldots,\alpha_{m},-\ell_3,n,\beta_1,\ldots\beta_{m-1}, \ell_n )
&=\Atree( \ell_n,\alpha_1,\ldots,\alpha_{m},-\ell_3,n,\beta_1,\ldots\beta_{m-1} )
\cr
&=-\hspace{-0.5cm}\sum_{\sigma\in OP\{\alpha\}\{\beta^T\cup n\}}
\hskip -.4 cm 
\Atree(  \ell_n,\sigma( \{\alpha\} \{ \beta^T\cup n\} ), -\ell_3)\,,
\cr}
\equn$$
so that
$$
\eqalign{
M^\tree(-\ell_3^- ,4^+,\ldots,n^+, \ell_n^-)
=
& -i\biggl[ \Atree(-\ell_3^- ,4^+,\ldots,n^+, \ell_n^-) 
\cr \null & \hskip -4.8 truecm 
\times \hskip -0.5truecm 
\sum_{\alpha\in {\cal S}_{m},\beta\in {\cal S}_{m-1} }
 f(\alpha)\bar f(\beta)
\hskip -.4 cm 
 \sum_{\sigma\in OP\{\alpha\}\{\beta^T\cup n\}}
\hskip -.4 cm 
\Atree(  \ell_n,\sigma( \{\alpha\} \{ \beta^T\cup n\} ), -\ell_3)
\biggr]
+{\cal P} (4,\ldots, n-1)\,.
\cr}
\equn$$
Using this we obtain
$$
\eqalign{
\hat c_{\NeqEight}^{(4^+,\ldots,n^+)1^-2^-3^-}
&= 2 
\hskip -0.5truecm 
\sum_{ \alpha\in {\cal S}_{m},\beta\in {\cal S}_{m-1}}
 f(\alpha)\bar f(\beta) 
\hskip -0.3truecm 
\sum_{\sigma\in OP\{\alpha\}\{\beta^T\cup n\}}
\hat c_{\NeqFour}^{(4^+,\ldots,n^+)1^-2^-3^-}
\hat c_{\NeqFour}^{(\sigma( \{\alpha\} \{ \beta^T\cup n\} ))3^-2^-1^-}
\cr
&\hspace{8cm}+{\cal P} (4,\ldots, n-1)\,.
\cr}
\equn$$
The explicit forms of these $\NeqFour$ coefficients may be found in 
refs.~\cite{Cachazo:2004dr,BDDK7,BDKn}.  The explicit form of
the gravity coefficient is then obtained by substituting these
into the expression.


\section{Twistor-Space Structure of Gravity Tree Amplitudes}
The twistor-space properties of supergravity amplitudes
are more complicated to analyse than the corresponding Yang-Mills
amplitudes since their support is on derivatives of
$\delta$-functions rather than simple $\delta$-functions. While the
MHV Yang-Mills amplitudes can be seen to have a simple $\delta$-function
support
$$\eqalign{
A_n^{\rm tree \;  MHV}&( 1^+,2^+,\ldots ,p^-,\ldots
,q^-,\ldots, n^+)(\lambda_i,\mu_i)\cr & \sim \int d^4 x
\prod_{i=1}^n\delta^2(\mu_{i \dot a}+x_{a\dot
a}\lambda_i^a)A_n^{\rm tree\; MHV}( 1^+,2^+,\ldots
,p^-,\ldots ,q^-,\ldots, n^+)(\lambda_i)\,,}
\equn
$$
where
$$
A_n^{\rm tree\; MHV}( 1^+,2^+,\ldots ,p^-,\ldots ,q^-,\ldots,
n^+) =  i { \spa{p}.q^4  \over \spa1.2\spa2.3\spa3.4 \cdots
\spa{n-1,}.n \spa{n}.1} \,,
\equn
$$
is the Parke-Taylor~\cite{ParkeTaylor} for MHV amplitudes. Expressions for the
gravity MHV amplitudes have been presented using the KLT relationship
together with factorisation by Berends, Giele and Kuijf~\cite{BerGiKu}.
Gravity MHV amplitudes will have a derivative of a $\delta$-function
support which mathematically can be expressed as
$$
\eqalign{
M_n^{\rm tree\; MHV}&
( 1^+,2^+,\ldots ,p^-,\ldots ,q^-,\ldots, n^+)(\lambda_i,\mu_i)
\cr
 & \hspace{5.5cm}\sim\int d^4 x
P\Big(-i\frac{\partial}{\partial\mu_{i \dot a}}\Big)
\prod_{i=1}^n
\delta^2(\mu_{i\dot a}+x_{a\dot a}\lambda_i^a)
\,,
\cr}
\equn
$$
following~\cite{WittenTopologicalString} where $P$ is a polynomial function. 
This has the geometric meaning that points are sitting
infinitesimally off lines in twistor space.

Consequently, we must test the twistor-space behaviour of gravity amplitudes
by acting multiple times with the $F_{ijk}$ and $K_{ijkl}$ operators
which we define as
$$
[F_{ijk} , \eta ] =
 \spa{i}.j \left[{ \partial\over \partial\tilde\lambda_k},\eta\right]
+\spa{j}.k \left[{ \partial\over
\partial\tilde\lambda_i},\eta\right] +\spa{k}.i \left[{
\partial\over
\partial\tilde\lambda_j},\eta\right]\,,
\equn
$$
and
$$
\eqalign{
K_{ijkl}=\frac14\Bigg[ \la {i j} \ra \epsilon^{ {\dot a
\dot b}} \frac{\partial}{\partial{\tilde \lambda}^{{\dot
a}}_{{k}}} \frac{\partial}{\partial {\tilde \lambda}^{{\dot
b}}_{{l}}} &-\la {i k} \ra {\epsilon}^{{\dot a \dot b}}
\frac{\partial}{\partial {\tilde \lambda}^{{\dot a}}_{{j}}}
\frac{\partial}{\partial {\tilde \lambda}^{{\dot b}}_{{l}}} +\la
{i l} \ra {\epsilon}^{{\dot a \dot b}} \frac{\partial}{\partial
{\tilde \lambda}^{{\dot a}}_{{j}}} \frac{\partial}{\partial
{\tilde \lambda}^{{\dot b}}_{{k}}}
\cr
&+\la {j k} \ra
{\epsilon}^{{\dot a \dot b}} \frac{\partial}{\partial {\tilde
\lambda}^{{\dot a}}_{{i}}} \frac{\partial}{\partial {\tilde
\lambda}^{{\dot b}}_{{l}}} +\la {j l} \ra {\epsilon}^{{\dot a \dot
b}} \frac{\partial}{\partial{\tilde \lambda}^{{\dot a}}_{{k}}}
\frac{\partial}{\partial {\tilde \lambda}^{{\dot b}}_{{i}}} -\la
{k l} \ra {\epsilon}^{{\dot a \dot b}} \frac{\partial}{\partial
{\tilde \lambda}^{{\dot a}}_{{j}}} \frac{\partial}{\partial
{\tilde \lambda}^{{\dot b}}_{{i}}} \Bigg]\,.}
\equn
$$

For Yang-Mills tree amplitudes the MHV amplitudes satisfy collinearity
conditions
$$
F_{ijk}  A^{\rm tree\;  MHV}_n (1.....n) = 0\,,
\equn$$
and the NMHV satisfy coplanarity relations
$$
K_{ijkl}  A^{\rm tree\;NMHV}_n (1.....n) = 0\,.
\equn$$

These conditions are manifest in the CSW
construction~\cite{Cachazo:2004kj} for gauge theory amplitudes.
Although, as yet, a similar construction does not
exist for gravity, one can determine the twistor-space properties by
direct computation.  Such an investigation may help shed light on the
origin of the difficulties encountered in finding MHV vertices for
gravity or finding a string theory dual. Although, Yang-Mills MHV tree
amplitudes are holomorphic (independent of $\tilde\lambda_{\dot a}$), the
KLT relationships imply that the gravity MHV tree amplitudes are
polynomial in $\tilde\lambda_{\dot a}$.  From the degree of the
polynomial we are guaranteed that
$$
F^P  M^{\rm tree\; MHV}_n (1.....n) = 0\,,
\hbox{\rm   \hskip 1. truecm  for  }\;\;\;  P >   2(n-3)\,.
\equn
$$
We now show that actually fewer powers of $F$ are required to
annihilate the tree amplitudes.  We also examine the coplanarity 
properties of the
NMHV amplitudes using the coplanar operator $K_{ijkl}$.

Consider first the five-point amplitude. There are two
inequivalent amplitudes: the MHV and the googly-MHV. These
satisfy~\cite{GBgravity}
$$
K^2  M_5^{\rm tree\; MHV}= 0  ,\,\,\,\,
KK'  M_5^{\rm tree\; MHV} \neq 0 ,\,\,\,\,
KK'K'' M_5^{\rm tree\; MHV }=0\,,
\equn
$$
and
$$
K^2  M^{\rm tree\; googly}_5 = KK'  M^{\rm tree\; googly}_5 =0\,,
\equn
$$
where $K$, $K'$ and $K''$ represent distinct $K_{ijkl}$.

By examining the tree amplitudes explicitly we have
$$
\eqalign{
F_{ijk}^{4}  M^{\rm tree\; MHV}_6  &  = 0\,,
\cr
F_{ijk}^{5}  M^{\rm tree\; MHV}_7  &  = 0\,,
\cr
F_{ijk}^{6}  M^{\rm tree\; MHV}_8  &  = 0\,,
\cr}
\equn
$$
and
$$
\eqalign{
K^3_{ijkl} M_6^{\rm tree\; (---+++)} & = 0\,,
\cr
K^4_{ijkl} M_7^{\rm tree\; (---++++)} & = 0\,.
\cr}
\equn
$$
These were checked by using computer algebra and by numerically
evaluating the expressions at arbitrary kinematic points.  The rapid
proliferation of terms in the gravity amplitudes as the number of
external legs increases makes further checks problematic.
In any case, this leads us to postulate the general behaviour,
$$
\eqalign{
F_{ijk}^{n-2}  M^{\rm tree\; MHV}_n  &  = 0\,,
\cr
K^{n-3}_{ijkl} M_n^{\rm tree\; NMHV} & = 0\,.
\cr}
\equn
$$


\section{Twistor-Space Structure of Gravity Box Coefficients}

In this section we show how the box coefficients of supergravity
one-loop amplitudes inherit a twistor-space structure directly from
box coefficients of super-Yang-Mills theory.  In particular, we show
that the box-coefficients for MHV gravity amplitudes have collinear support
whilst the box-coefficients of NMHV gravity amplitudes have coplanar support,
similar to the situation for gauge theory.

Unitarity links the tree amplitudes to the imaginary parts
of loop amplitudes. For example considering the cut in a one-loop
amplitude we have
$$
\eqalign{
C_{i,\ldots,j}   & \equiv
{ i \over 2} \int \dlips\biggl[ M^{\rm tree}(\ell_1,i,i+1,\ldots,
j,\ell_2)  \cr & \hspace{5.cm}\times
M^{\rm tree}(-\ell_2,j+1,j+2,\ldots,i-1,-\ell_1)
\biggr]
\cr
& = {\rm Im}_{K_{i,\ldots,j} > 0}  M^{\rm 1\hbox{-}loop}
=
\sum_a   c_a  \; {\rm Im}_{K_{i,\ldots,j} > 0}( F_a)\,,
\cr}
\equn
$$
where the one-loop amplitude is expressed as a sum of integral
functions $F_a$ multiplied by rational functions $c_a$.  One can
use this expression, and more generalised unitarity expressions to
deduce information on the behaviour of the $c_a$.  Specifically,
consider the action of a differential twistor space operator
${\cal O}$ which satisfies
$$
{\cal O}M^{\rm tree}(\ell_1,i,i+1,\ldots,j,\ell_2)=0\,,
\equn
$$
and where ${\cal O}$ only depends on legs ${i,\ldots,j}$.  Naively, the
action of ${\cal O}$ on the cut gives zero however due to the
``holomorphic anomaly''~\cite{Cachazo:2004by,Cachazo:2004dr} the
action of ${\cal O}$ produces a $\delta$-function within the
integral of the cut leading to~\cite{BBKR,Cachazo:2004dr}
$$
{\cal O}C_{i,\ldots,j} = {\rm rational }\,,
\equn
$$
after the integral has been performed.  For the case where $F_a$ is a
box integral function the imaginary parts are logarithms of the momentum
invariants. In general,
$$
{\cal O}C_{i,\ldots,j} ={\cal O}\biggl(
\sum_a c_a  \; {\rm Im}_{K_{i,\ldots,j} > 0}( F_a) \biggr)\,,
\equn
$$
can only be satisfied provided~\cite{Cachazo:2004dr}
$$
{\cal O}c_a  =0\,.
\equn
$$
We shall apply operators of the form ${\cal O}=F^n$ and ${\cal O}=K^n$ 
to the box integral coefficients, 
using generalised unitarity as a guide to the 
expected properties.

The first example we will look at is a box-coefficient of the
five-point MHV one-loop amplitude. (The action of $F_{ijk}$
does not depend on how the helicities are assigned since the
arrangement of helicities will only change a holomorphic factor
in the box coefficient.) Consider the various cuts of 
the box,

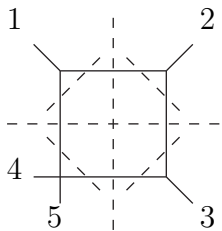
\begin{center}
\begin{picture}(100,100)(0,0)
\Line(30,30)(30,70)
\Line(70,30)(70,70)
\Line(30,30)(70,30)
\Line(70,70)(30,70)
\Line(30,30)(20,30)
\Line(30,30)(30,20)
\Line(70,30)(80,20)
\Line(70,70)(80,80)
\Line(30,70)(20,80)
\Text(10,90)[l]{$1$}
\Text(83,90)[l]{$2$}
\Text(83,15)[l]{$3$}
\Text(25,15)[l]{$5$}
\Text(10,33)[l]{$4$}
\DashLine(50,10)(50,90){4}
\DashLine(10,50)(90,50){4}
\DashLine(25,45)(45,25){4}
\DashLine(25,55)(45,75){4}
\DashLine(55,25)(75,45){4}
\DashLine(75,55)(55,75){4}
\end{picture}
\end{center}

\def\ce{c_{\NeqEight}}

Consider first the cut $C_{123}$, where the gravity MHV five-point
tree amplitude $M^\tree(\ell_5,1,2,3,\ell_3)$ is isolated on one side
of the cut.  This tree is annihilated by $F_{123}^3$. Hence we
conclude that the same property should hold for the coefficient:
$F_{123}^3 \ce^{(45)123}=0$.  Similarly by examining the cuts
$C_{451}$ and $C_{345}$, which isolate legs $4,5,1$ and $3,4,5$
respectively, we deduce that
$F^3_{145}\ce^{(45)123}=F^3_{345}\ce^{(45)123}=0$.  For the remaining
choices of $F_{ijk}$ we must consider more generalised cuts. For
example in the case of $F_{124}$ we can consider the cut $C_{4512}$.
By analytically continuing to signature $(--++)$ such cuts are
possible and will be non-vanishing~\cite{BrittoUnitarity} and allow us
to deduce information on the coefficients. In this case the gravity
tree amplitude $M^\tree(\ell_3,4,5,1,2,\ell_2)$ is a six-point MHV
tree, annihilated by, {\it e.g.}, $F_{124}^4$ and we deduce
$F^4_{124}\ce^{(45)123}=0$. Summarising we have
$$
F_{123}^3 \ce^{(45)123}=F^3_{145}\ce^{(45)123}=F^3_{345}\ce^{(45)123}=0,
 \;\;\;\;\;
F_{ijk}^4 \ce^{(45)123}=0  \;\;\forall \; i,j,k\,.
\equn
$$
A similar conclusion holds for all other box coefficients in the five
point amplitude, simply by permuting the legs.

Considering the six-point MHV-amplitude there are two types of
boxes to consider: the one mass and two-mass-easy boxes,
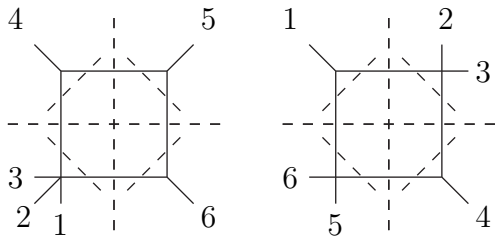
\begin{center}
\begin{picture}(100,100)(0,0)
\Line(30,30)(30,70)
\Line(70,30)(70,70)
\Line(30,30)(70,30)
\Line(70,70)(30,70)
\Line(30,30)(20,30)
\Line(30,30)(30,20)
\Line(70,30)(80,20)
\Line(70,70)(80,80)
\Line(30,70)(20,80)
\Line(20,20)(30,30)
\Text(10,90)[l]{$4$}
\Text(83,90)[l]{$5$}
\Text(83,15)[l]{$6$}
\Text(27,12)[l]{$1$}
\Text(13,15)[l]{$2$}
\Text(10,30)[l]{$3$}
\DashLine(50,10)(50,90){4}
\DashLine(10,50)(90,50){4}
\DashLine(50,10)(50,90){4}
\DashLine(10,50)(90,50){4}
\DashLine(25,45)(45,25){4}
\DashLine(25,55)(45,75){4}
\DashLine(55,25)(75,45){4}
\DashLine(75,55)(55,75){4}

\end{picture}
\begin{picture}(100,100)(0,0)
\Line(30,30)(30,70)
\Line(70,30)(70,70)
\Line(30,30)(70,30)
\Line(70,70)(30,70)
\Line(30,30)(20,30)
\Line(30,30)(30,20)
\Line(70,30)(80,20)
\Line(70,70)(70,80)
\Line(70,70)(80,70)
\Line(30,70)(20,80)
\Text(10,90)[l]{$1$}
\Text(69,90)[l]{$2$}
\Text(83,70)[l]{$3$}
\Text(83,15)[l]{$4$}
\Text(27,12)[l]{$5$}
\Text(10,30)[l]{$6$}
\DashLine(50,10)(50,90){4}
\DashLine(10,50)(90,50){4}
\DashLine(50,10)(50,90){4}
\DashLine(10,50)(90,50){4}
\DashLine(25,45)(45,25){4}
\DashLine(25,55)(45,75){4}
\DashLine(55,25)(75,45){4}
\DashLine(75,55)(55,75){4}
\end{picture}
\end{center}
and examining the various cuts as before we find
$$
\hspace{0.89cm}
\eqalign{
F^3_{ijk}\ce^{\rm 1m} &= 0\,,  \hskip .5 cm  \{i,j,k\}=\{1,2,3\}, \{4,5,6\}\,,
\cr
F^4_{ijk}\ce^{\rm 1m} &= 0\,, \hskip .5 cm 
{\rm if} \;\;\;\{i,j,k\}\in \{3,4,5,6\} \;\;\; 
{ \rm or} \;\;\;\{i,j,k\}\in \{1,4,5,6\},
\;\;\;
\cr
F^5_{ijk}\ce^{\rm 1m} &= 0\,,
\hskip .5 cm 
\forall \{i,j,k\}\,,
\cr}
\equn
$$
as well as
$$
\hspace{1.42cm}
\eqalign{
F^3_{ijk}\ce^{{\rm 2m}e} &= 0\,, \hskip .5 cm 
 \{i,j,k\}=\{1,2,3\}, \{4,5,6,\}, \{1,5,6\}, \{2,3,4\}\,,
\cr
F^4_{ijk}\ce^{{\rm 2m}e} &= 0\,, \hskip .5 cm {\rm if}
\{i,j,k\}\in \{1,2,3,4\}  \;\;\;  { \rm or}     \;\;\;  \{i,j,k\}\in \{1,4,5,6\}\,,
\cr
F^5_{ijk}\ce^{{\rm 2m}e} &= 0\,, \hskip .5 cm 
\forall \{i,j,k\}\,,
\cr}
\equn
$$
where $\ce^{\rm 1m}$ and $\ce^{{\rm 2m} e}$ are shorthands for the
one-mass and two-mass-easy box coefficients $\ce^{(123)456}$ and
$\ce^{1(23)4(56)}$ As we can see, the box-coefficients have 
``derivative of $\delta$-function`` collinear support although the degree of
annihilation by $F^n$ depends on the choice of indices.  As before
the properties of other coefficients appearing in the amplitude can 
be obtained simply by permuting labels.

Continuing in this way, by inspecting the general $n$-point case, 
we can predict
$$
F^{n-1}_{ijk} \ce^{n\hbox{-}{\rm point}} = 0 , \;\;\;\;   \forall i,j,k\,,
\equn
$$
indicating collinearity in twistor space.
However, in some cases the value of $n-1$ is not optimal in the
sense that smaller multiples of $F_{ijk}$ will suffice for some
values of $\{i,j,k\}$.

We now consider the action of $K^N$ on the box-coefficients of
the NHMV amplitude $M(1^-,2^-,3^-,4^+,5^+)$ (which is the parity dual
of a MHV amplitude). To act with $K_{ijkl}^N$, we must use a cut
which isolates at least four external legs. Specifically, to
analyse the behaviour of $K_{2345}^N$ we can examine the
$C_{2345}$ cut, where one of the tree amplitudes is
$M^\tree(\ell_1^h,2^-,3^-,4^+,5^+,\ell_5^{h'})$. If we only consider this cut
then the values of $h$ and $h'$ may both be negative and we have a
tree amplitude with four negative helicities: such amplitudes do
not have coplanar support.  However if we consider the quadruple
generalised cut then the only possible non-vanishing solution is

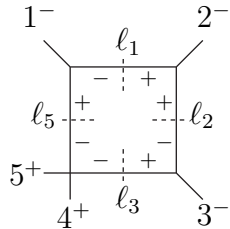
\begin{center}
\begin{picture}(100,100)(0,0)
\DashLine(50,73)(50,61){2}
\DashLine(50,39)(50,27){2}
\DashLine(27,50)(39,50){2}
\DashLine(61,50)(73,50){2}

\Line(30,30)(30,70)
\Line(70,30)(70,70)
\Line(30,30)(70,30)
\Line(70,70)(30,70)

\Line(30,30)(30,20)
\Line(30,30)(20,30)

\Line(30,70)(20,80)
\Line(70,30)(80,20)
\Line(70,70)(80,80)

\Text(25,14)[l]{$4^+$}
\Text(78,15)[l]{$3^-$}
\Text(78,90)[l]{$2^-$}
\Text(12,90)[l]{$1^-$}

\Text(7,30)[l]{$5^+$}

\Text(35,40)[c]{$^-$}
\Text(35,55)[c]{$^+$}
\Text(42,64)[c]{$^-$}
\Text(60,64)[c]{$^+$}
\Text(66,55)[c]{$^+$}
\Text(66,40)[c]{$^-$}
\Text(60,33)[c]{$^+$}
\Text(42,33)[c]{$^-$}

\Text(20,52)[c]{$\ell_5$}
\Text(52,20)[c]{$\ell_3$}
\Text(80,52)[c]{$\ell_2$}
\Text(52,80)[c]{$\ell_1$}
\end{picture}
\end{center}
indicating that the problem helicity does not contribute to
this box coefficient.  The possible tree amplitude
$M^\tree(\ell_1^+,2^-,3^-,4^+,5^+,\ell_5^-)$ is indeed annihilated by
$K_{2345}^3$ which thus requires $K^3_{2345}\ce=0$. Explicit
computation confirms that $K^3_{ijkl}\ce=0$ for this box-coefficient.

Continuing in this way we can deduce that at least,
$$
K^{n-2} \ce^{n\hbox{-}{\rm point}}=0\,,
\equn
$$
although in some cases we need to apply less powers of $K$.  We can
consider a generic box with three massless legs, {\it e.g.}, with exactly
one negative helicity on each legs and at least three legs
attached to each massive vertex,

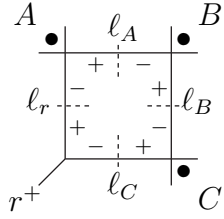
\begin{center}
\begin{picture}(100,100)(0,0)
\DashLine(50,73)(50,61){2}
\DashLine(50,39)(50,27){2}
\DashLine(27,50)(39,50){2}
\DashLine(61,50)(73,50){2}

\Line(30,30)(30,70)
\Line(70,30)(70,70)
\Line(30,30)(70,30)
\Line(70,70)(30,70)

\Line(30,30)(20,20)

\Line(30,70)(20,70)
\Line(30,70)(30,80)

\Line(70,30)(70,20)
\Line(70,30)(80,30)

\Line(70,70)(70,80)
\Line(70,70)(80,70)

\Text(15,15)[c]{$r^+$}
\Text(85,15)[c]{$C$}
\Text(85,85)[c]{$B$}
\Text(15,85)[c]{$A$}
\Text(75,25)[c]{$\bullet$}
\Text(75,75)[c]{$\bullet$}
\Text(25,75)[c]{$\bullet$}

\Text(35,40)[c]{$^+$}
\Text(35,55)[c]{$^-$}
\Text(42,64)[c]{$^+$}
\Text(60,64)[c]{$^-$}
\Text(66,55)[c]{$^+$}
\Text(66,40)[c]{$^-$}
\Text(60,33)[c]{$^+$}
\Text(42,33)[c]{$^-$}
\Text(20,52)[c]{$\ell_r$}
\Text(52,20)[c]{$\ell_C$}
\Text(80,52)[c]{$\ell_B$}
\Text(52,80)[c]{$\ell_A$}
\end{picture}
\end{center}
where $A$, $B$ and $C$ are sets of indices of external lines
including a single negative helicity. Immediately we can
deduce that
$$\eqalign{
F^{N}_{ijk} \ce =0 \; , \;\; \{i,j,k\} \in A\cup \{ r\} ,\;\;\;\;\;\;\;\;\;\;\;
F^{M}_{ijk} \ce &=0 \; , \;\; \{i,j,k\} \in B ,\cr
F^{P}_{ijk} \ce &=0 \; , \;\; \{i,j,k\} \in C\cup \{ r\} \,,
\cr}\equn
$$
where $N$, $M$ and $P$ are integers depending on the number of
legs attached to the corner, and
$$
K^{S}_{ijkl} \ce =0 \;\; \{i,j,k,l\} \in A\cup B ,\hspace{1cm}
K^{T}_{ijkl} \ce =0 \;\; \{i,j,k,l\} \in B\cup C  \,.
\equn
$$

\FIGURE[h]{
\begin{picture}(250,150)(-30,0)
\SetWidth{1.5}

\Line(20,10)(160,130)
\Line(80,120)(190,10)

\Line(00,30)(200,46)

\Text(50,24)[c]{$r^+$}
\Text(110,22)[c]{$C$}
\Text(67,77)[c]{$A$}
\Text(152,80)[c]{$B$}

\Vertex(48,34){4}

\Vertex(137.5,41){4}
\Vertex(100,38){4}
\Vertex(80,36.5){4}

\Vertex(135,65){4}
\Vertex(120,80){4}
\Vertex(150,50){4}

\Vertex(100,78.5){4}
\Vertex(82,63){4}
\Vertex(65,48.5){4}
\end{picture}
\label{NMHVTwistorFigure}
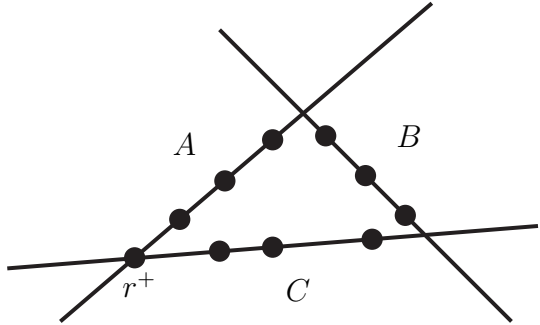
\caption{The three-mass box coefficients of the NMHV one-loop amplitude have 
derivative of $\delta$-function support in twistor space on three
intersecting lines lying in a plane.  This
diagram is identical to the one found in ref.~\cite{BDKn} for the
NMHV amplitudes of $\NeqFour$ super-Yang-Mills theory.}
}

\FIGURE[h]{
\hskip 6 cm 
\begin{picture}(250,110)(0,0)
\DashLine(50,73)(50,61){2}
\DashLine(50,39)(50,27){2}
\DashLine(27,50)(39,50){2}
\DashLine(61,50)(73,50){2}

\Line(30,30)(30,70)
\Line(70,30)(70,70)
\Line(30,30)(70,30)
\Line(70,70)(30,70)

\Line(30,70)(20,70)
\Line(30,70)(30,80)

\Line(70,30)(70,20)
\Line(70,30)(80,30)

\Line(70,70)(70,80)
\Line(70,70)(80,70)

\Line(30,20)(30,30)
\Line(20,30)(30,30)

\Text(15,15)[c]{$D$}
\Text(25,25)[c]{$\bullet$}
\Text(85,15)[c]{$C$}
\Text(85,85)[c]{$B$}
\Text(15,85)[c]{$A$}
\Text(75,25)[c]{$\bullet$}
\Text(75,75)[c]{$\bullet$}
\Text(25,75)[c]{$\bullet$}

\Text(35,40)[c]{$^+$}
\Text(35,55)[c]{$^-$}
\Text(42,64)[c]{$^+$}
\Text(60,64)[c]{$^-$}
\Text(66,55)[c]{$^+$}
\Text(66,40)[c]{$^-$}
\Text(60,33)[c]{$^+$}
\Text(42,33)[c]{$^-$}
\Text(20,52)[c]{$\ell_D$}
\Text(52,20)[c]{$\ell_C$}
\Text(80,52)[c]{$\ell_B$}
\Text(52,80)[c]{$\ell_A$}
\end{picture}
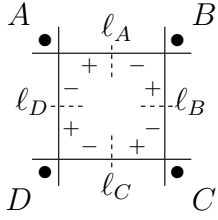
\caption{A four-mass box with clusters of legs indicated by $A,B,C,D$.
The dots represent arbitrary numbers of external legs.\label{FourMassBoxABCDFigure}}
}
%

The important point here is that the supergravity three-mass
box-coefficient in twistor space has a topology inherited from that of
super-Yang-Mills: the points lie on three intersecting lines with two
of these lines intersecting at point $r$, as shown in
\fig{NMHVTwistorFigure}.  This picture is identical to that for
super-Yang-Mills~\cite{BDKn}--essentially because the argument leading
to it is identical albeit with the important difference that we must
act with multiple copies of $F$ and $K$.
This example illustrates how the twistor
picture for $\NeqEight$ box-coefficients will be inherited from that
of $\NeqFour$ super-Yang-Mills.

As a check of the predicted coplanarity for the $\NeqEight$
supergravity box coefficients, we have explicitly computed the action
of the $K^n$ operators on the six-point box coefficients. Using
computer algebra, and numerically evaluating the results at a generic
kinematic point we have verified that all gravity six-point
coefficients have a derivative of a $\delta$-function support on
planes in twistor space, confirming the predicted patterns.

\FIGURE{
\begin{picture}(240,170)(-25,0)
\SetWidth{1.5}

\Line(-10,15)(160,85)
\Line(-12,25)(200,46)
\Line(200,10)(150,165)
\Line(140,60)(160,160)
\Text(90,20)[c]{$D$}
\Text(70,80)[c]{$A$}
\Text(190,95)[c]{$C$}
\Text(130,120)[c]{$B$}

\Vertex(137.5,40){4}
\Vertex(100,36){4}
\Vertex(60,32){4}

\Vertex(183.7,60){4}
\Vertex(179,75){4}
\Vertex(172.4,95){4}
\Vertex(164.4,120){4}

\Vertex(146.5,94){4}
\Vertex(149.5,109){4}
\Vertex(152.6,124){4}

\Vertex(120,69){4}
\Vertex(100,60.5){4}
\Vertex(82,53){4}
\Vertex(65,46){4}
\end{picture}
\label{NNMHVTwistorFigure}
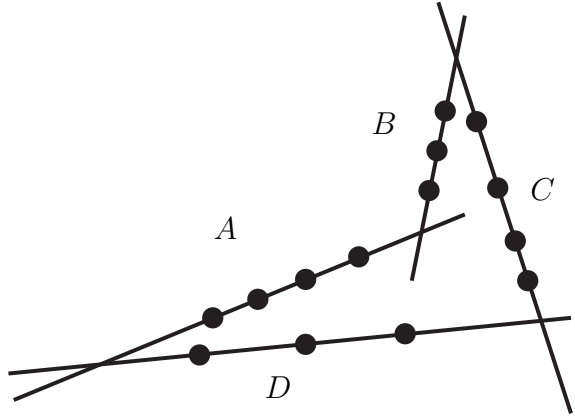
\caption{The four-mass box-coefficients of N$^2$MHV amplitudes have
derivative of $\delta$-function support in twistor
space on four intersecting lines.}
}

The twistor support of the four-mass box-coefficients of the N$^2$MHV
gravity amplitudes is also very similar to the Yang-Mills N$^2$MHV
case~\cite{BDKn}.
If we consider the generic four-mass box with clusters of legs
$A$, $B$, $C$ and $D$ as indicated in \fig{FourMassBoxABCDFigure},
for the N$^2$MHV case the clusters form four MHV trees.
From this observation we obtain,
$$
\eqalign{
F^r_{ijk} \ce &= 0  , \; \{ijk\} \in A\;\;\;
F^{r'}_{ijk}\ce = 0  , \;  \{ijk\} \in B\,,
\cr
F^{r''}_{ijk}\ce &= 0  , \;  \{ijk\} \in C\;\;\;
F^{r'''}_{ijk}\ce = 0  , \;  \{ijk\} \in D\,,
\cr}\equn
$$
and
$$
\eqalign{\hspace{1cm}
K^s_{ijk}\ce &= 0  , \; \{ijk\} \in A\cup B\,, \hskip 1 cm
K^{s'}_{ijk}\ce = 0  , \;  \{ijk\} \in B\cup C\,,
\hspace{1cm}\cr
K^{s''}_{ijk}\ce &= 0  , \;  \{ijk\} \in C\cup D\,, \hskip 1 cm
K^{s'''}_{ijk}\ce = 0  , \;  \{ijk\} \in D\cup A\,,
\cr}\equn
$$
which gives us a picture of four pair-wise intersecting lines or
of points lying on a pair of intersecting planes shown in
\fig{NNMHVTwistorFigure}.  Again this matches the picture for
$\NeqFour$ super-Yang-Mills theory~\cite{BDKn}, except that
the lines have derivative of $\delta$-function support.

In summary, the box-coefficients of one-loop amplitudes in $\NeqEight$
supergravity have a twistor-space structure inherited from $\NeqFour$
super-Yang-Mills amplitudes.


\section{Conclusions}

Gauge and gravity theories are two of the cornerstones of modern
theoretical physics. Explicit calculations within these theories have
been very fruitful for uncovering and testing theoretical properties.
In the gauge theory case, such calculations are also crucial for
comparisons of theory against experiments.

In this paper we have investigated the twistor-space properties of
both tree and loop amplitudes in $\NeqEight$ supergravity
which is especially interesting because of its close
connection to $D=11$ supergravity, which again is closely linked to $M$-theory.
It is also believed to be the gravity theory with the best ultraviolet
properties: the first potential divergence occurs at no less than
five loops~\cite{BDDPR,Howe:2002ui}.

In general, gravity tree amplitudes are simple to obtain via the low
energy limit of the KLT relationship~\cite{KLT,BerGiKu,GravityReview} between
open and closed strings.  In the case of loop amplitudes the
technically simplest theory to deal with is that of $\NeqEight$
supergravity.  Here we have computed the coefficients of the box
functions in $\NeqEight$ supergravity in order to determine their
twistor-space properties.

Loop calculations in quantum gravity theories are notoriously
difficult: direct calculation using Feynman diagram techniques being
significantly more difficult than the equivalent gauge theory ones.
In this paper we followed the logic of
refs.~\cite{BDDPR,Bern:1998xc,Bern:1998sv}, making use of the
unitarity method together with the KLT relations to obtain
supergravity loop amplitudes from gauge theory tree amplitudes.
The observation of ref.~\cite{BrittoUnitarity} that quadruple
cuts freeze the loop integrals helped simplify our discussion.

We have produced sample formul\ae{} for supergravity box
integral coefficients from known gauge theory amplitudes.  Following
similar logic, it should be possible to obtain the full $n$-point
gravity one-loop amplitudes by recycling the equivalent Yang-Mills
gauge theory amplitudes.  Specifically we have calculated box integral
coefficients in $\NeqEight$ supergravity using the corresponding ones
of $\NeqFour$ super-Yang-Mills theory. In many ways these box
coefficients are simpler objects than tree amplitudes.  As has
happened for gauge
theory~\cite{BDDK7,BDKn,Britto:new,Roiban:2004ix,WittenRecursive}, it
seems likely, although perverse, that loop amplitudes will prove a
route to simplifying gravity tree amplitudes.

Factorisation properties suggest that only box integral functions
appear in one-loop $\NeqEight$ supergravity amplitudes.  If this is
true, one can obtain complete one-loop $\NeqEight$ supergravity
amplitudes simply by evaluating the box integral coefficients.  This,
however, would require a non-trivial cancellation, since the gravity
amplitudes are {\it a priori} not cut-constructible from the
four-dimensional tree amplitudes.  Rather surprisingly, 
it also seems to imply that at one-loop $\NeqEight$
supergravity has a power counting identical to that of $\NeqFour$
super-Yang-Mills theory.  It would be very interesting to prove
whether box function are the only integral functions that appear in
$\NeqEight$ one-loop amplitudes.  Assuming a proof is found, it would
also be very interesting to investigate possible implications of these
types of cancellations on the higher-loop ultraviolet divergences of
$\NeqEight$ supergravity.

We also showed how twistor space properties of gauge theory
amplitudes are inherited by gravity loop amplitudes, via the Kawai-Lewellen
and Tye tree relations and the unitarity method.  The relatively simple
twistor-space structure of the gravity amplitudes described in this
paper hints that there may be a twistor-space string theory
interpretation.  We hope that further investigations will provide new
insight into quantum gravity.

\begin{acknowledgments}
We thank Lance Dixon, Harald Ita and David Kosower for many useful
discussions. This research was supported in part by the US Department
of Energy under grant DE--FG03--91ER40662 and in part by the PPARC.
We also thank Academic
Technology Services at UCLA for computer support on a 48 CPU Beowulf
cluster, where computations of the twistor-space structure of the
gravity amplitudes were performed.

\end{acknowledgments}

\vfill\eject

\end{document}